\documentclass[%
 reprint,
 amsmath,amssymb,
 aps,superscriptaddress,
nobibnotes,prb,
floatfix,
longbibliography]{revtex4-1}
\usepackage{graphicx}
\usepackage{dcolumn}
\usepackage{float}
\usepackage[mathscr]{euscript}
\usepackage{xcolor}
\usepackage[linkbordercolor=green]{hyperref}
\usepackage{tikz}
\usepackage{tabularx} 
\hypersetup{
    colorlinks=true,
    linkcolor=red,
    filecolor=magenta,      
    urlcolor=cyan,
    citecolor = red,
}

\usepackage{amsmath}
\usepackage[normalem]{ulem}

\newcommand{\bb}[1]{\boldsymbol{#1}}

\newcommand*\bbi[2]{\text{$[{#1} {#2}]_i$}}
\newcommand*\bi[2]{\text{$[j_{#1} j_{#2}]_i$}}

 \newcommand{\clcir}{
    \begin{tikzpicture}
        \draw [thick, fill=black] (0,0) circle (0.075); 
    \end{tikzpicture}
}

\newcommand{\rect}{
    \begin{tikzpicture}
        \draw[thick, fill=black] (0,0.0) rectangle (0.05,0.25); 
    \end{tikzpicture}
}
\begin{document}

\title{Micelle Forming
Linear-Dendritic Block Copolymers: A Theoretical Comparison
between Random Hyperbranched and 
Precise Dendrimer Polymer Architectures}

\author{Marios Giannakou}
\email{mgiannak@uni-mainz.de}
 \affiliation{Institut für Physik, Johannes Gutenberg-Universität, Mainz, 55099, Germany}
 
\author{Oleg Borisov}%
\email{oleg.borisov@univ-pau.fr}
\affiliation{%
Institut des Sciences Analytiques et de
Physico-Chimie pour l’Environnement et les Matériaux, Pau, 64053, France}%
\author{Friederike Schmid}
 \email{friederike.schmid@uni-mainz.de}
 \affiliation{Institut für Physik, Johannes Gutenberg-Universität, Mainz, 55099, Germany}

\preprint{APS/123-QED}

\begin{abstract}
\textbf{ABSTRACT:} 
 Hyperbranched block copolymers offer a simpler and more efficient synthesis route compared to more traditional dendritic systems, while still providing exceptional control over surface functionality and self-assembly. This makes them ideal candidates for engineering nanoparticles with tailored properties for applications such as drug delivery and sensing. Here we use self-consistent field calculations to compare the micelle structures formed by copolymers with a polydisperse hyperbranched (LHBC), monodisperse dendritic (LDBC), and linear solvophilic blocks. Representative LHBC structures were generated by molecular dynamics simulations mimicking the slow-monomer addition protocol. We find that LHBC micelles are more stable, have a lower critical micelle concentration, and are better at accommodating larger drug payloads than LDBC micelles, and these properties further improve with increasing polydispersity. LHBC micelles also offer more terminal ends for functionalization than LDBC micelles for LDBCs with up to four branching generations, with the number of terminal ends being surprisingly independent of the LHBC polydispersity. Our findings highlight the superiority of LHBC micelles in flexibility and performance over LDBC micelles.  
\end{abstract}
\maketitle
\section{\label{sec:intro}Introduction}
\begin{figure}
    \centering
    \includegraphics[width=0.48\textwidth]{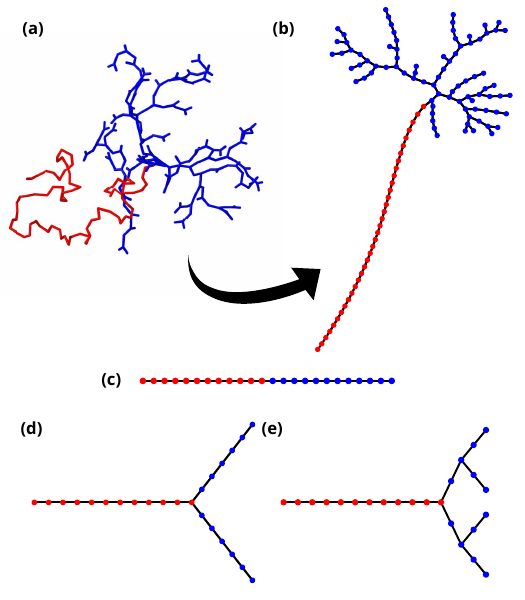}
    \caption{Examples of molecular structures. 
    Solvophobic part is shown in red and solvophilic part in blue. (a) Example of an LHBC molecule produced from a molecular dynamics simulation (see Section \ref{sec:mdmodel}). (b) Graph representation of the polymer molecule in (a) where 
    each filled circle represents a monomer. Note that the maximum number of generations in this particular example is ten. (c) Representation of a symmetric linear diblock molecule. (d,e) Representation of a LDBC molecule of generation one (d) and generation two (e).}
    \label{fig:mol_sketch}
\end{figure}

Block copolymers have seen a long and sustained interest, both in experiments and theory, mainly due to their ability to self-assemble into a variety of nanoassemblies. This ability stems from the fact that the constituting blocks of the polymers are made from different types of often incompatible monomers that would like to demix, however due to the connectivity of the blocks they instead microphase separate \cite{hamley1998physics}. In the melt regime, for example, even the simplest type of block copolymer, the linear diblock polymer, can self-assemble into a variety of periodic structures like lamellar, hexagonal, spherical, gyroid and more\cite{hamley1998physics,Matsen1994,Khandpur1995,Mai2012,Bates2019}, with a periodicity determined largely by the macromolecular weight of the molecules themselves and thereby in the nanoscale range\cite{Meier1969,AnalChimActa1986,hamley1998physics}. Such a capability is highly desired in a range of applications, such as surface pattering\cite{Park1997}, thin films\cite{Thurn-Albrecht2000,Krausch2002}, filtration \cite{Ahn2014} and many more\cite{Lazzari2003,Kim2010}. On the other hand, if a solvent is present that is selective toward one of the types of blocks, but poor toward the others, then the polymers may self-assemble into a variety of states depending on the concentration of the polymers, the molecular weight of the polymer, and other parameters \cite{Lodge2002}. Some common examples include spherical micelles, elongated micelles, worm-like micelles, or vesicles\cite{Hanley2000,Battaglia2006,Mai2012,Barnhill2015}. Such structures have been intensely investigated and have a wide range of applications, {\it e.g.}, in solubilization\cite{Gadelle1995}, stabilization\cite{Sakai2004}, as nanoreactors\cite{Peters2012}, for drug encapsulation and delivery\cite{Kataoka2012,Bose2021} and many others\cite{Hamley2005}.
In the present article we focus on polymeric
micelles, which hold promise as nanocarriers for encapsulating and
transporting drugs.  Micelles do
this by incorporating the often hydrophobic drug\cite{Bose2021} into
their cores, thus solubilizing and protecting it from
the highly complex environment {\it in
vivo}\cite{Kim2010}. In addition, as drugs need to
circulate in the body for some time to reach their target sites, it is vital that
the drug release from the nanocarrier happens over hours and not immediately \cite{Kim2010} after administration. After entering the bloodstream, the nanoparticles find themselves in a highly dilute environment, much below the critical micelle concentration
(CMC), whereupon they disassemble quickly and thus
release their drug payload. Polymeric micelles, on the other
hand, have a relatively
low critical micelle concentration, which enhances their stability and slows down their
disassembly to a large extent\cite{Owen2012}. Moreover, rather than passively
delivering drugs to a site, a more selective strategy involves actively targeting
the sites by releasing the drug payload near or inside the affected
cells. In this regard, polymeric micelles offer a variety of possibilities. For example, by
introducing stimuli-responsive functional groups or monomers, it is possible to induce the release of a drug at a specific
site using triggers such as light, temperature or pH \cite{Kelley2013}. Lastly,
decorating micelles with specific moieties, such
as ligands,
enables targeting of desired sites that have specific
receptors for said ligand\cite{Allen2002,Ruiz2022}, thereby minimizing the contact with healthy cells.
Thus, polymeric micelles that serve as drug delivery vehicles
should combine a variety of attributes. Fortunately, the vast array of synthetic
protocols\cite{Hadjichristidis2005} has made
it possible to construct a variety of exotic polymers. One such class, that combines multiple benefits 
and has attracted considerable interest in recent
years, is linear dendritic block copolymers (LDBCs). These polymers consist of a linear
solvophobic block and a precise branched structure consisting of
hydrophilic blocks, resembling a
tree\cite{Gitsov2008,Wurm2011,Fan2016}. In solvent,
they self-assemble into an even greater variety of structures
than linear block copolymers \cite{Whitton2015,Liu2019}. Additionally, LDBCs offer several other
advantages over linear block copolymers, including smaller  micelle sizes, lower
aggregation numbers and a greater number of chain ends available
for functionalization\cite{Lebedeva2018}. However, synthesizing LDBCs with a precise branch
structure -- {\it i.e.}, with controlled macromolecular weight and
branch generations -- requires a
multipot process\cite{Tomalia1985,Grayson2001}.
This complexity results in relatively high production costs compared
to simpler copolymers. 
An alternative approach which has gained popularity in recent years is to use their less precise cousins, the so-called linear hyperbranched block copolymers (LHBCs) \cite{Nuhn2013,oikawa2013one}. In contrast to the case of LDBCs, the branched component of LHBCs is
highly random. This randomness arises from their
synthetic protocols, which are both blessings and a curse. For
example LHBCs can be synthesized in a
one-pot process\cite{Cuneo2020}, considerably reducing production
complexity. However, this simplification often comes at the cost of high macromolecular weight
and topological polydispersity\cite{Uhrich1992,Kim1990}. As drug delivery
vehicles must be monodisperse in size and exhibit similar physiological characteristics between batches, it is important for the polymers
to form well-defined structures \cite{Andresen2020}. High
macromolecular weight polydispersity can lead to
undesirable assemblies\cite{Schmitt2012}. To address
this issue, methods that reduce polydispersity, such as
slow-monomer addition \cite{Barriau2005}, have been developed. It should be noted that a certain low degree of macromolecular weight polydispersity may have a positive effect on  micelle size uniformity, as has been demonstrated for linear block copolymers\cite{Mantha2019}.  
Theoretical studies on micelle formation have mostly focused on
monodisperse linear block copolymers
\cite{DEGENNES1978,Noolandi1983,Leibler1983,Leermakers1995,
Nelson1997,zhulina_theory_2012} and LDBCs\cite{Wang2013, Lebedeva2018,lebedeva_self-assembly_2019,Brito2023}. 
A few simulation studies have investigated micelle self-assembly and morphological transitions in solutions of hyperbranched copolymers with irregular architectures \cite{Tan2015,Tan2017,Tan2019, Hao2020}; however, the systems  were still monodisperse in the sense that all molecules were identical. Only few
studies have considered effects of molecular weight polydispersity
\cite{Gao1993, Linse1994, Lynd2008, Doncom2017,
Mantha2019,Giannakou2024}, and the effects of topological polydispersity remain largely unexplored.  
Here, we attempt to elucidate some of the properties of micelles composed of
polydisperse LHBCs, and compare them with their counterparts made of
monodisperse linear diblock copolymers and LDBCs. Schematic
pictures of such polymers are
shown in Fig.\ \ref{fig:mol_sketch}b-e. In the case of LDBCs, the solvophobic blocks comprising the dendritic part have the same total number of monomers and the
number of terminal ends doubles with each generation.
We investigate a range of metrics such as the
morphologies of the micelles, the terminal end
distributions, the stability of micelles, their CMC values, and their encapsulation capacities for a model solvophobic drug molecule. We also investigated the limiting molecular weight polydispersity that can still be tolerated. To this end, we employ molecular dynamics (MD) simulations to model the slow-monomer addition method \cite{sunder1999controlled,Barriau2005,schuell2013polydispersity} and construct
a variety of LHBCs with predetermined macromolecular length (weight) polydispersity. The molecular architectures are then extracted and the
self-assembly of the molecules is evaluated in the grand canonical
ensemble using the Self-Consistent Field Theory (SCFT) framework\cite{Schmid1998}. 
\section{\label{sec:methods}Model and Methods}
\subsection{Molecular Dynamics Model}\label{sec:mdmodel}
\begin{figure}
     \centering
     \includegraphics[width=0.25
     \textwidth]{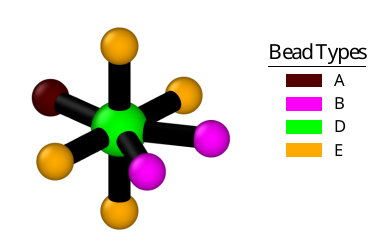}
     \caption{Sketch of the AB$_2$ monomer used in the MD simulations. Each color corresponds to a different type of bead as indicated, and bonds are depicted in black. In the SCFT calculations, this whole monomer is turned into a single solvophilic segment.}
     \label{fig:monomer_md}
 \end{figure}

We employed MD simulations to mimic the slow-monomer addition protocol \cite{rabbel2013statistical} and used beads labeled A to F, to represent various components of the LHBC. Beads F and C represent polystyrene and the macroinitiator respectively, while the rest are used to represent the AB$_2$ monomers and are configured in a star-like fashion as shown in Fig. \ref{fig:monomer_md}: The center bead of the star (type D)
is connected to two beads B and one bead A such that the four of them
form a Y-shape, and further inert beads E are added to stabilize this
structure. Beads A can interact via an attractive
potential with beads C and B, simulating the irreversible conjugation
of $\mathrm{AB_{2}}$ monomers with the macroinitiator and with each other.

The detailed interactions between each bead type are outlined below:
\begin{eqnarray}\label{eq:mdpotentials}
&&U^{\text{harm.}}_{ij}=\frac{1}{2}k_{h}(r_{ij}-r_{o})^2 
 \\  \nonumber
&&\mbox{for bonded beads 
  \,F-F,\ F-C,\ C-C,\ A-D,\ B-D,\ D-E,}
\\
&&U^{\text{LJ}}_{ij}=4\epsilon
  \left[ \left(\frac{\sigma}{r_{ij}}\right)^{12}
    -\left(\frac{\sigma}{r_{ij}}\right)^{6} \right] 
\\ \nonumber
&&\mbox{for all bead pairs except \,A-C,\,A-B},
\\
&&U^{\text{bind.}}_{ij}=-d \,
  \cos\left(\frac{r_{ij}\pi}{2 r_{c}}\right)
  \Theta(r_{c}-r_{ij})
\\ \nonumber
&&\mbox{for bead  pairs \, A-C,\, A-B,}
\\
&&U^{\text{cos}}_{ijk}=\frac{1}{2}k_{c}(\cos
  \left(\theta_{ijk}\right)-\cos\left(\theta_{o}\right))^2
  \\ \nonumber
&&\mbox{for bonded bead triplets\,  A-D-B,\ 
 B-D-B},
 \\
&&U^{\text{dihedral}}_{ijkl}
   =\frac{1}{2}k_d(1-\cos\left(2\phi_{ijkl}\right))^2
  \\ \nonumber
&&\mbox{for beads E-E-E-E in the same star unit.} \nonumber
\end{eqnarray}
Here $r_{ij}$ denotes the distance between beads $i$ and $j$,
$\theta_{ijk}$ is the angle between particles $i$, $j$ and $k$,
 $\phi_{ijkl}$ is the angle between the two planes
formed by beads $i$, $j$, $k$ and beads $j$, $k$, $l$ respectively, and $\Theta$ refers to the Heaviside function 
($\Theta(x)=1$ for $x>0$, $\Theta(x)=0$ otherwise).
The parameters are $\kappa_{h}=100\epsilon$, $\rho_{o}=\sigma$, 
$d=100\epsilon$, $r_{c}=\sigma/2$, $\kappa_{c}=100\epsilon$, 
$\theta_{o}=5\pi/6$ for A-D-B and 
$\theta_{o}=2\pi/3$ for B-D-B.

We simulated the aforementioned system under constant temperature and volume
conditions using Langevin dynamics as implemented in the HOOMD-blue
molecular dynamics package \cite{Anderson2020}. Starting with a linear chain of $84$ F-beads followed by $8$ C-beads
connected in a sequential arrangement, we then introduced a designated
number of $\mathrm{AB_{2}}$ monomers. This number is sampled randomly
from a Schulz-Zimm distribution\cite{zimm1948apparatus} with an
average value of $\Bar{N}_{\mathrm{AB_2}} =$ $76$ monomers, and varying, but prescribed polydispersity
index $\text{PDI}= \overline{N^2_{\mathrm{AB_2}}}/\Bar{N}_{\mathrm{AB_2}}^2$. We note that the choice of $\Bar{N}_{\mathrm{AB_2}} =$ $76$ is based on the fact that in SCF, F beads act as solvophobic monomers, while C beads and AB$_2$ monomers act as identical solvophilic monomers, thus the resulting LHBCs are, on average, symmetric in terms of the solvophobic-to-solvophilic monomer ratio. For the case of PDI = $1$, the ensemble generated consists of LHBCs that are monodisperse in length yet display a diversity of topologies.

The $\mathrm{AB_{2}}$ monomers are added
sequentially, with the condition that the preceding monomer must first
be attached to the growing central molecule before a new monomer can
be introduced. This prevents premature
connections between free $\mathrm{AB_{2}}$ monomers. A schematic representation of such a polymer molecule and
its graph structure is shown in
Fig.\ \ref{fig:mol_sketch}a,b.
The graph representation of this molecule, along
with others that constitute the polydisperse
ensemble of LHBCs, is subsequently recorded and used
for further calculations within the SCF
framework. To avoid confusion we note that, although more than two types of MD monomers are introduced in the construction of the LHBC polymers, the MD monomers are then mapped onto only two types of segments, either solvophobic or solvophilic, in the SCFT model.

The SCF calculations are done in batches B$1$-B$4$, consisting of $128$ different polymers each, which are a result of the "greedy algorithm". This algorithm sorts the $512$ polymers, which we refer to as the BA batch, into four equally sized sub-batches (B$1$-B$4$). It does this by progressively filling
these sub-batches while tracking the total sum of monomers in each
batch. It then assigns the next polymer to the sub-batch with the
lowest total, ensuring that no sub-batch exceeds the target of $128$
polymers.
More details about the SCF simulations are provided in
Section \ref{sec:scfmodel}.

\subsection{SCFT Model}\label{sec:scfmodel}

To model a system of copolymers with solvophobic (H) and
solvophilic (P) monomers in solvent (S), capable of
exchanging polymer chains with its environment (bath), we employ SCFT  
calculations in the grand canonical ensemble.

We consider a polymer solution in
implicit solvent, modeled according to the Sanchez–Lacombe 
theory\cite{sanchez-lacombe,Schmid1998}, 
and characterize the system in terms of spatially varying monomer
volume fractions $\phi_H(\bb{r})$ and $\phi_P(\bb{r})$ that depend
on the corresponding monomer number densities $\rho_{\alpha}(\bb{r})$
and the monomeric volumes $v_{\alpha}=v_{P}$ via
$\phi_{\alpha}=\rho_{\alpha} v_{\alpha}$.  Thus the solvent volume
fraction is given by $\phi_S(\bb{r}) = 1 - \phi_H(\bb{r}) -
\phi_P(\bb{r})$ and the solvent number density is given by $\rho_S=\phi_S/v_S$,
where $v_S$ is the volume of a solvent molecule. The grand canonical
free energy is given by \cite{Schmid1998}:
\begin{eqnarray}\label{eq:free_energy}
   &&\beta F_{\text{GC}}=
     \left(U_{\text{inter.}}
       - \frac{1}{v^{*}}
       \int \text{d}\bb{r}\sum_{\alpha}^{H,P}
         (\rho_{\alpha}v^{*})W_{\alpha}\right.
         \nonumber\\
      &&\left.\quad\quad\quad\quad\quad\quad
        -\sum_{i}^{n_{T}}\exp{(\beta\mu_{i})}Q_{i}\right)
        \nonumber\\ 
   &&U_{\text{inter.}}=\frac{1}{v^{*}} 
      \left( \int \text{d}\bb{r} \, 
      \sum_{\alpha}^{H,P}
        \chi_{\alpha S} \: \phi_{\alpha}(\bb{r}) \phi_{S}(\bb{r})
        \right. 
      \\
   &&\quad\quad\quad\quad\quad
     +\frac{1}{2}\sum_{\alpha,\beta}^{H,P}
        \chi_{\alpha\beta}
       \: \phi_{\alpha}(\bb{r}) \: \phi_{\beta}(\bb{r})
        \nonumber\\
   &&\quad\quad\quad\quad\quad+ \:
      v^* (\: 
       \rho_S(\bb{r})
        \ln (\phi_{S}\big(\bb{r}))
      -\rho_S(\bb{r})\big) \bigg)\nonumber,
\end{eqnarray}
where $v^{*}$ is a reference volume, $U_{\text{inter.}}$ is the
interaction potential which also includes the translational
entropy of the solvent molecules, $\chi_{\alpha\beta}$ are the Flory-Huggins parameters between species $\alpha$ and $\beta$, $W_{\alpha}$ are the
self-consistent fields, ${\mu_{i}}$ and ${Q_{i}}$ is the chemical
potential and the single chain partition functions of chains of
type ${i}$ respectively, ${n_{T}}$ is the number of different
types of polymers, and $V$ is the volume of the 
system. 

In these grand canonical SCF calculations, we assume the polymers in
the micelle to be in chemical equilibrium with a homogeneous solution
of chains of type $i$ with global average polymer volume fraction
$\Bar{\phi}$. The chemical potentials $\mu_i$ are then given by:
\begin{equation}\label{eq:chemicalpotential}
    {\exp(\beta \mu_{i}
      +\ln}(\Bar{N}))=\frac{w_{i}\Bar{\phi}V}{\Bar{Q_{i}}v_{P}},
\end{equation}
where ${\Bar{Q_{i}}}$ is the single chain partition function of chain type
$i$ in the homogeneous state and ${w_{i}}$ is the fraction of chains of type $i$ in the bath such that ${\sum_{i}^{n_{T}}w_{i}=1}$. Also,
${\Bar{N}=\sum_{i}^{n_{T}}w_{i}N_{i}}$ is the average chain length and $N_{i}$ is the length of polymer type $i$. The derivation of
Eq.\ \eqref{eq:chemicalpotential} is given in the Appendix \ref{addcalc}. 

In our study, we consider copolymers  that are
separated into blocks, each consisting exclusively of either solvophobic or solvophilic monomers. We categorize the blocks into three
 groups based on their connectivity: (1) Stem (SM,
one per molecule), (2) Internal (IL), and (3) Terminal
(TL). Stem and terminal blocks each have one free end, while internal blocks have none. Blocks are delimited by junctions, which
encompass both the internal branch points and free ends.  For each
molecule type $i$, the junctions are numbered consecutively, starting
from zero, which is assigned to the free end of the stem block. Thus,
a given block in a chain of type $i$ can be identified by the pair
\bi{1}{2}
of confining junctions. Moreover, we assign orientations
to molecules, defining the forward direction as running from the stem
to the terminal blocks. An
example illustrating the nomenclature is given in Fig.\ \ref{fig:cartoon}. 

\tikzset{every picture/.style={line width=0.75pt}} 
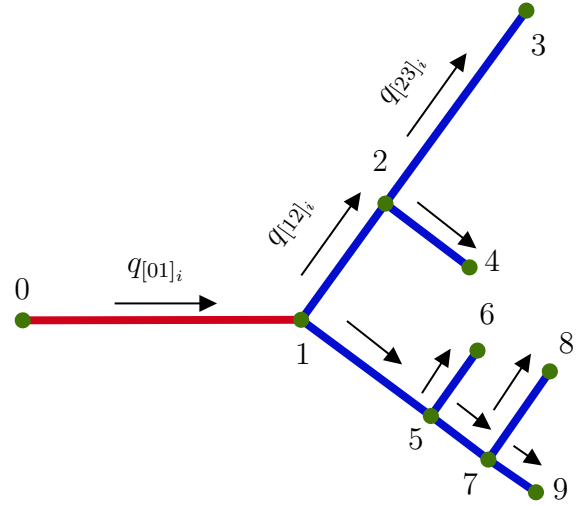
\begin{figure}
    \centering
\begin{tikzpicture}[x=0.75pt,y=0.75pt,yscale=-1,xscale=1]
\draw [color={rgb, 255:red, 2; green, 13; blue, 208 }  ,draw opacity=1 ][line width=3]    (415,255) -- (444,277) ;
\draw [color={rgb, 255:red, 208; green, 2; blue, 27 }  ,draw opacity=1 ][line width=3]    (209.17,206.73) -- (349.58,206.46) ;
\draw [color={rgb, 255:red, 2; green, 13; blue, 208 }  ,draw opacity=1 ][line width=3]    (349.58,206.46) -- (392.07,147.81) ;
\draw [color={rgb, 255:red, 2; green, 13; blue, 208 }  ,draw opacity=1 ][line width=3]    (392.07,147.81) -- (463.17,50.39) ;
\draw    (255.5,198.19) -- (304.17,198.19) ;
\draw [shift={(307.17,198.19)}, rotate = 180] [fill={rgb, 255:red, 0; green, 0; blue, 0 }  ][line width=0.08]  [draw opacity=0] (8.93,-4.29) -- (0,0) -- (8.93,4.29) -- cycle    ;
\draw    (349.66,184.06) -- (377.94,144.45) ;
\draw [shift={(379.68,142.01)}, rotate = 125.52] [fill={rgb, 255:red, 0; green, 0; blue, 0 }  ][line width=0.08]  [draw opacity=0] (8.93,-4.29) -- (0,0) -- (8.93,4.29) -- cycle    ;
\draw    (402.99,113.9) -- (431.27,74.28) ;
\draw [shift={(433.01,71.84)}, rotate = 125.52] [fill={rgb, 255:red, 0; green, 0; blue, 0 }  ][line width=0.08]  [draw opacity=0] (8.93,-4.29) -- (0,0) -- (8.93,4.29) -- cycle    ;
\draw [color={rgb, 255:red, 2; green, 13; blue, 208 }  ,draw opacity=1 ][line width=3]    (392.07,147.81) -- (434.5,180) ;
\draw [color={rgb, 255:red, 2; green, 13; blue, 208 }  ,draw opacity=1 ][line width=3]    (349.58,206.46) -- (415,255) ;
\draw [color={rgb, 255:red, 2; green, 13; blue, 208 }  ,draw opacity=1 ][line width=3]    (415,255) -- (438.5,222) ;
\draw [color={rgb, 255:red, 2; green, 13; blue, 208 }  ,draw opacity=1 ][line width=3]    (444,277) -- (475,232.5) ;
\draw [color={rgb, 255:red, 2; green, 13; blue, 208 }  ,draw opacity=1 ][line width=3]    (444,277) -- (468,293.5) ;
\draw  [color={rgb, 255:red, 65; green, 117; blue, 5 }  ,draw opacity=1 ][fill={rgb, 255:red, 65; green, 117; blue, 5 }  ,fill opacity=1 ] (411.47,254.5) .. controls (411.75,252.55) and (413.55,251.2) .. (415.5,251.47) .. controls (417.45,251.75) and (418.8,253.55) .. (418.53,255.5) .. controls (418.25,257.45) and (416.45,258.8) .. (414.5,258.53) .. controls (412.55,258.25) and (411.2,256.45) .. (411.47,254.5) -- cycle ;
\draw  [color={rgb, 255:red, 65; green, 117; blue, 5 }  ,draw opacity=1 ][fill={rgb, 255:red, 65; green, 117; blue, 5 }  ,fill opacity=1 ] (440.47,276.5) .. controls (440.75,274.55) and (442.55,273.2) .. (444.5,273.47) .. controls (446.45,273.75) and (447.8,275.55) .. (447.53,277.5) .. controls (447.25,279.45) and (445.45,280.8) .. (443.5,280.53) .. controls (441.55,280.25) and (440.2,278.45) .. (440.47,276.5) -- cycle ;
\draw  [color={rgb, 255:red, 65; green, 117; blue, 5 }  ,draw opacity=1 ][fill={rgb, 255:red, 65; green, 117; blue, 5 }  ,fill opacity=1 ] (346.06,205.96) .. controls (346.33,204.01) and (348.14,202.66) .. (350.08,202.93) .. controls (352.03,203.21) and (353.39,205.01) .. (353.11,206.96) .. controls (352.83,208.91) and (351.03,210.26) .. (349.08,209.98) .. controls (347.14,209.71) and (345.78,207.91) .. (346.06,205.96) -- cycle ;
\draw  [color={rgb, 255:red, 65; green, 117; blue, 5 }  ,draw opacity=1 ][fill={rgb, 255:red, 65; green, 117; blue, 5 }  ,fill opacity=1 ] (388.54,147.31) .. controls (388.82,145.36) and (390.62,144) .. (392.57,144.28) .. controls (394.51,144.56) and (395.87,146.36) .. (395.59,148.31) .. controls (395.32,150.25) and (393.51,151.61) .. (391.57,151.33) .. controls (389.62,151.06) and (388.26,149.25) .. (388.54,147.31) -- cycle ;
\draw  [color={rgb, 255:red, 65; green, 117; blue, 5 }  ,draw opacity=1 ][fill={rgb, 255:red, 65; green, 117; blue, 5 }  ,fill opacity=1 ] (205.64,206.23) .. controls (205.92,204.28) and (207.72,202.93) .. (209.67,203.2) .. controls (211.61,203.48) and (212.97,205.28) .. (212.69,207.23) .. controls (212.42,209.17) and (210.61,210.53) .. (208.67,210.25) .. controls (206.72,209.98) and (205.36,208.18) .. (205.64,206.23) -- cycle ;
\draw  [color={rgb, 255:red, 65; green, 117; blue, 5 }  ,draw opacity=1 ][fill={rgb, 255:red, 65; green, 117; blue, 5 }  ,fill opacity=1 ] (434.97,221.5) .. controls (435.25,219.55) and (437.05,218.2) .. (439,218.47) .. controls (440.95,218.75) and (442.3,220.55) .. (442.03,222.5) .. controls (441.75,224.45) and (439.95,225.8) .. (438,225.53) .. controls (436.05,225.25) and (434.7,223.45) .. (434.97,221.5) -- cycle ;
\draw  [color={rgb, 255:red, 65; green, 117; blue, 5 }  ,draw opacity=1 ][fill={rgb, 255:red, 65; green, 117; blue, 5 }  ,fill opacity=1 ] (471.47,232) .. controls (471.75,230.05) and (473.55,228.7) .. (475.5,228.97) .. controls (477.45,229.25) and (478.8,231.05) .. (478.53,233) .. controls (478.25,234.95) and (476.45,236.3) .. (474.5,236.03) .. controls (472.55,235.75) and (471.2,233.95) .. (471.47,232) -- cycle ;
\draw  [color={rgb, 255:red, 65; green, 117; blue, 5 }  ,draw opacity=1 ][fill={rgb, 255:red, 65; green, 117; blue, 5 }  ,fill opacity=1 ] (464.47,293) .. controls (464.75,291.05) and (466.55,289.7) .. (468.5,289.97) .. controls (470.45,290.25) and (471.8,292.05) .. (471.53,294) .. controls (471.25,295.95) and (469.45,297.3) .. (467.5,297.03) .. controls (465.55,296.75) and (464.2,294.95) .. (464.47,293) -- cycle ;
\draw  [color={rgb, 255:red, 65; green, 117; blue, 5 }  ,draw opacity=1 ][fill={rgb, 255:red, 65; green, 117; blue, 5 }  ,fill opacity=1 ] (430.97,179.5) .. controls (431.25,177.55) and (433.05,176.2) .. (435,176.47) .. controls (436.95,176.75) and (438.3,178.55) .. (438.03,180.5) .. controls (437.75,182.45) and (435.95,183.8) .. (434,183.53) .. controls (432.05,183.25) and (430.7,181.45) .. (430.97,179.5) -- cycle ;
\draw  [color={rgb, 255:red, 65; green, 117; blue, 5 }  ,draw opacity=1 ][fill={rgb, 255:red, 65; green, 117; blue, 5 }  ,fill opacity=1 ] (459.64,49.89) .. controls (459.92,47.95) and (461.72,46.59) .. (463.67,46.87) .. controls (465.61,47.14) and (466.97,48.95) .. (466.69,50.89) .. controls (466.42,52.84) and (464.61,54.2) .. (462.67,53.92) .. controls (460.72,53.64) and (459.36,51.84) .. (459.64,49.89) -- cycle ;
\draw    (408.16,147.06) -- (435.64,168.65) ;
\draw [shift={(438,170.5)}, rotate = 218.15] [fill={rgb, 255:red, 0; green, 0; blue, 0 }  ][line width=0.08]  [draw opacity=0] (8.93,-4.29) -- (0,0) -- (8.93,4.29) -- cycle    ;
\draw    (372.66,207.06) -- (398.14,227.14) ;
\draw [shift={(400.5,229)}, rotate = 218.24] [fill={rgb, 255:red, 0; green, 0; blue, 0 }  ][line width=0.08]  [draw opacity=0] (8.93,-4.29) -- (0,0) -- (8.93,4.29) -- cycle    ;
\draw    (428.16,248.56) -- (443.13,260.16) ;
\draw [shift={(445.5,262)}, rotate = 217.78] [fill={rgb, 255:red, 0; green, 0; blue, 0 }  ][line width=0.08]  [draw opacity=0] (8.93,-4.29) -- (0,0) -- (8.93,4.29) -- cycle    ;
\draw    (446.66,252.06) -- (463.85,226) ;
\draw [shift={(465.5,223.5)}, rotate = 123.41] [fill={rgb, 255:red, 0; green, 0; blue, 0 }  ][line width=0.08]  [draw opacity=0] (8.93,-4.29) -- (0,0) -- (8.93,4.29) -- cycle    ;
\draw    (456.66,270.06) -- (468.53,278.29) ;
\draw [shift={(471,280)}, rotate = 214.73] [fill={rgb, 255:red, 0; green, 0; blue, 0 }  ][line width=0.08]  [draw opacity=0] (8.93,-4.29) -- (0,0) -- (8.93,4.29) -- cycle    ;
\draw    (410.5,243.5) -- (423.37,223.52) ;
\draw [shift={(425,221)}, rotate = 122.8] [fill={rgb, 255:red, 0; green, 0; blue, 0 }  ][line width=0.08]  [draw opacity=0] (8.93,-4.29) -- (0,0) -- (8.93,4.29) -- cycle    ;

\draw (385.87,94.34) node [anchor=north west][inner sep=0.75pt]  [rotate=-305.56] [align=left][font=\large] 
{$\displaystyle q_{\bbi{2}{3}}$};
\draw (329.37,165.84) node [anchor=north west][inner sep=0.75pt][font=\large] [rotate=-305.56] [align=left] 
{$\displaystyle q_{\bbi{1}{2}}$};
\draw (260.28,174.13) node [anchor=north west][inner
sep=0.75pt][font=\large]  [align=left] 
{$\displaystyle  q_{\bbi{0}{1}}$};
\draw (203.5,184) node [anchor=north west][inner sep=0.75pt][font=\large]   [align=left] {0};
\draw (345,217) node [anchor=north west][inner sep=0.75pt]   [align=left][font=\large] {1};
\draw (384,119.5) node [anchor=north west][inner sep=0.75pt]   [align=left][font=\large] {2};
\draw (464,61) node [anchor=north west][inner sep=0.75pt]   [align=left][font=\large] {3};
\draw (440.5,170) node [anchor=north west][inner sep=0.75pt][font=\large]   [align=left] {4};
\draw (438,197) node [anchor=north west][inner sep=0.75pt]
[align=left][font=\large] {6};
\draw (402,260) node [anchor=north west][inner sep=0.75pt]
[align=left][font=\large] {5};
\draw (429.5,282.5) node [anchor=north west][inner sep=0.75pt][font=\large]   [align=left] {7};
\draw (478,211.5) node [anchor=north west][inner sep=0.75pt][font=\large]   [align=left] {8};
\draw (475,285) node [anchor=north west][inner sep=0.75pt][font=\large]   [align=left] {9};
\end{tikzpicture}
    \caption{Cartoon representation of a hyperbranched polymer
    indexed to $i$. Red indicates that block $j$ is solvophobic while
    blue indicates that the block is solvophilic. Green indicates the
    junction points numbered here from $0$ to $9$. For clarity, only some
    of the forward propagators are shown.}
    \label{fig:cartoon}
\end{figure}
For each block $\bi{1}{2}$, we calculate a forward propagator
$q_{\bi{1}{2}}(\bb{r},s)$ and a backward propagator
$q^{\dagger}_{\bi{1}{2}}(\bb{r},s)$, where $s = n/\Bar{N}$ and
$n$ is a monomer count. This is done by solving the
modified diffusion equations\cite{Grason2005}
\begin{align}\label{eq:mde}
    \frac{\partial q_{\bi{1}{2}} 
       (\bb{r},s)}
      {{\partial}s}
      &=\left(\Bar{N}\frac{b^2}{6}\nabla^{2}
        -\Bar{N} W_{\bi{1}{2}} 
         (\bb{r}) \right)
         q_{\bi{1}{2}} (\bb{r},s) 
         \\ \nonumber
    \frac{\partial q^{\dagger}_{\bi{1}{2}}(\bb{r},s)}
        {{\partial}s}
       &=-\left(\Bar{N}\frac{b^2}{6}\nabla^{2}-\Bar{N}
         W_{\bi{1}{2}} (\bb{r})\right)
         q^{\dagger}_{\bi{1}{2}}(\bb{r},s),
\end{align}
where we assumed the statistical segment length of the monomers $b$ to 
be the same throughout the polymer. Here, $W_{\bi{1}{2}}$ is either ${W_{H}}$ 
or ${W_{P}}$, depending on the monomer type of block
$\bi{1}{2}$. Eqs.\ \eqref{eq:mde} are
solved for values of $s$ in the interval
${s\in[0,s^{\text{max}}_{\bi{1}{2}}]}$, where $\Bar{N}
s^{\text{max}}_{\bi{1}{2}}$ is the macromolecular length of block
$\bi{1}{2}$, with initial conditions given by the following relations\cite{Grason2005}:
 \begin{align}\label{eq:in0}
q_{\bi{1}{2}} (\bb{r},0)
       &= 1
    \quad \text{for} \quad
       \bi{1}{2} = \bbi{0}{1} \in \text{SM}
\end{align}
\begin{align}\label{eq:in1}
q_{\bi{1}{2}} (\bb{r},0)
      &= q_{\bi{3}{1}} (\bb{r},s^{\text{max}}_{\bi{3}{1}})
         \: q^{\dagger}_{\bi{1}{4}}(\bb{r},0)
        \nonumber\\[10pt]
 & \hspace{-1cm} \text{for} \quad \bi{1}{2} \notin \text{SM} \quad \text{and},
        \nonumber\\[5pt]
 & \hspace{-2cm} \bi{3}{1} \in \text{SM or IL},
           \; \bi{1}{4} \in \text{IL or TL}
\end{align}
\begin{align}\label{eq:in2}
q^{\dagger}_{\bi{1}{2}}(\bb{r},s^{\text{max}}_{\bi{1}{2}})  
  &= 1 
  \quad \text{for} \quad
   \bi{1}{2} \in \text{TL}
\end{align}
\begin{align}\label{eq:in3}
q^{\dagger}_{\bi{1}{2}}(\bb{r},s^{\text{max}}_{\bi{1}{2}})
    &= q^{\dagger}_{\bi{2}{3}}(\bb{r},0) \:
       q^{\dagger}_{\bi{2}{4}}(\bb{r},0)
    \nonumber\\[10pt]
 & \hspace{-2cm} \text{for} \quad
    \bi{1}{2} \notin \text{TL},
    \; \bi{2}{3}, \: \bi{2}{4} \in \text{IL or TL.}
\end{align}
Eqs.\ \eqref{eq:in0} and \eqref{eq:in2} are the initial conditions for the free ends of the polymer, while Eqs. \eqref{eq:in1} and \eqref{eq:in3} are the initial conditions for the inner junction points of the polymer. For example, in Fig.\ \ref{fig:cartoon}, junction points $1,\,2,\,5$ and $7$ are inner junction points while the rest are free ends.
Based on these initial conditions, we first
calculate the backward propagators of the chain, starting from
the terminal groups and proceeding "backward" along the chain, up
until the stem's backward propagator is calculated. Then we
repeat the procedure for the forward propagator in
the reverse order. Once the propagators have been calculated, the
volume fractions can be determined {\it via} the following expression:
\begin{align}\label{eq:volfra}
        &\phi_{\alpha}(\bb{r})
          =\sum_{i}^{n_{T}}\phi_{\alpha,i}(\bb{r})
          =\sum_{i}^{n_{T}}\exp(\beta \mu_{i}
          + \ln(\Bar{N})) \\[3pt]
    &\times \sum_{\bi{1}{2}}
       \int_{0}^{s^{\text{max}}_{\bi{1}{2}}}
       \,\text{d} s \,
       q_{\bi{1}{2}} (\bb{r},s) 
        q^{\dagger}_{\bi{1}{2}} 
        (\bb{r},s)
       \theta_{\alpha,\bi{1}{2}}
       ,
      \nonumber
\end{align}
where ${\phi_{\alpha,i}(\bb{r})}$ is the volume fraction
contribution from chain $i$ to monomer type ${\alpha}$ and
$\theta_{\alpha,\bi{1}{2}}$ is one if the block
 $\bi{1}{2}$ is  of type ${\alpha}$, and zero otherwise.
The single chain partition function of chain of type $i$
can be evaluated from the backward propagators of the 
corresponding stem blocks,
\begin{equation}
Q_i = \int \text{d} \bb{r} \: q^{\dagger}_{\bbi{0}{1}}(\bb{r},0).
\end{equation}
Finally, to close the self-consistent loop,
the fields $W_\alpha(\bb{r})$ are calculated
from the functional derivatives of $U_{\text{inter.}}$ with 
respect to the monomeric {\it number} densities\cite{Schmid1998},
$\rho_\alpha = \phi_\alpha/v_\alpha$, as:
\begin{align}\label{eq:field}
    W_{\alpha}(\bb{r})&
      =\frac{\delta U_{\text{inter.}}[\phi]} 
        {\delta \rho_\alpha(\bb{r})}  
 \nonumber\\
    &=\frac{v_{\alpha}}{v^{*}}
    \big( \chi_{\alpha S} \phi_S 
        + \sum_{\beta}^{H,P} (\chi_{\alpha \beta} - \chi_{\beta S})
         \phi_\beta
        - \frac{v^*}{v_S} \ln (\phi_S)\big). 
\end{align}
Given an initial field ${W_{\alpha}}$, we solve Eqs.\
\eqref{eq:mde}, calculate new volume fractions using Eq.
\eqref{eq:volfra}, calculate new fields using Eq.\ \eqref{eq:field}, 
mix the new fields with the old ones using lambda mixing\cite{muller_fschmid}, and repeat 
the loop until the following convergence
criterion is reached:
\begin{equation}\label{eq:conve}
    \text{CF}=\sum_{\alpha}^{H,P}
      \int \text{d}\bb{r} \, 
       (\phi_{\alpha}^{\text{new}}-\phi_{\alpha}^{\text{old}})^{2}
       <10^{-12}.
\end{equation}
All SCF calculations were performed with
periodic boundary conditions in a simulation box of volume
${V=15\times15\times15}\,[\Bar{R}_{g}^{3}]$, using ${1024}$,
${128\times128}$ and ${64\times64\times64}$ grid points for one, two
and three dimensional simulations respectively. 
The rest of the parameters were chosen as
${\chi_{HP}}\Bar{N}=30$,
${\chi_{HS}}\Bar{N}=61$,
${\chi_{PS}}\Bar{N}=27$, 
$v_S/(v^*\bar{N})=0.02$ and $v^{*}=v_{{P}}$ such that
the equilibrium morphology in a system of symmetric
diblock copolymers is a spherical micelle. To accelerate the numerical computation of the propagators for highly symmetric architectures like LDBCs, we implemented schemes similar to those in Yong and Kim \cite{yong2025dynamic}, which avoid redundant calculations of identical propagators.
\section{\label{sec:results}Results and Discussion}
In this section, we first examine the results related to the size and
topological polydispersity of LHBCs generated from MD simulations. We
then present results from SCF calculations, comparing micelles formed
by polydisperse ensembles of LHBCs with those formed by monodisperse
ensembles of linear diblocks or LDBCs of various generations. Key experimentally 
relevant quantities such as the critical micelle
concentrations (CMC), the equilibrium
morphologies, the volume fraction profiles, the
terminal end distributions, the number of chains ${n_{M}}$ and
terminal ends ${c_{M}}$ per micelle, the micelle size
distributions, and the energy penalty associated with asphericity, are
discussed. Finally, we investigate the drug encapsulation capacity of
these micelles by evaluating the encapsulation of solvophobic
homopolymers. For a fair comparison, we limit the study to systems with
solvophobic-to-solvophilic monomer ratio maintained at $1:1$
for all monodisperse LDBC and linear diblock systems, and on average, at $1:1$ for the
polydisperse LHBC systems. We note that, in some
of the following plots, we refer to the linear diblock chain as zeroth generation LDBC. Below, lengths
are mostly given in units of the average radius of gyration $\Bar{R}_g = b \sqrt{\Bar{N}/6}$ and the free energy $F$ will be given in units of $k_B T= \beta^{-1}$ and
rescaled with the Ginzburg parameter $\Bar{C} = \Bar{R}_g^3/v^*
\Bar{N}$.

\subsection{Generation of representative LHBC polymer sets}\label{sec:generation}
As noted in the introduction, slow monomer addition can yield polymers
with low macromolecular length polydispersity, which shows
particular promise for applications. Therefore, we focus on LHBCs
synthesized by using this approach. Specifically, we modeled the slow-monomer addition
protocol outlined in Barriau \textit{et al}\cite{Barriau2005}. In
this process, a linear polystyrene block is initially conjugated to a
short, linear hydroxylated polybutadiene block, which serves as a
macroinitiator for the subsequent gradual addition of glycidol,
ultimately forming the LHBC molecule. Here, glycerol acts as an
$\mathrm{AB_{2}}$-type monomer, thus the branching
points in the resulting hyperbranched polyglycerol have a degree of
three.

To model the slow monomer
addition protocol, we used single-chain coarse-grained MD simulations. In such simulations, a linear block was conjugated to a linear macroinitiator and a predetermined number of
$\mathrm{AB_{2}}$ monomers, that can irreversibly bond with the macroinitiator and other $\mathrm{AB_{2}}$ monomers, were added sequentially to the growing molecule. This predetermined random number followed a Schulz-Zimm distribution with a specific polydispersity index (PDI) and average number (see Section \ref{sec:mdmodel}.)

We investigated different values of PDI, and for each PDI, we simulated the creation of $512$ independent polymers. To keep the
SCF simulations manageable and enable assessing statistical
errors in the SCF results, we divided the complete batch (BA) into four
subbatches (B$1$, B$2$, B$3$, B$4$) of $128$ polymers each. Instead of randomly
selecting polymers from the BA batch, we used a "greedy number
partitioning" algorithm to assign polymers to sub-batches (see Section \ref{sec:mdmodel}).

\begin{figure}
    \centering
    \includegraphics[width=0.48\textwidth]{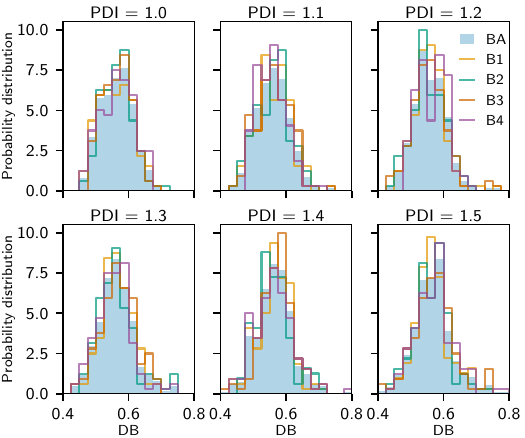}
    \caption{Probability distributions of the degree of branching (DB)
    for the LHBCs with different values of polydispersity index PDI. The whole colored
    distribution represents the complete batch (BA), while the other
    four colors represent the sub-batches (B$1$-B$4$). Note that the DB
    values are calculated only for the solvophilic part of the polymer.
    \label{fig:db_distr}}
\end{figure}

This method ensures that the four sub-batches have similar
average chain lengths and was also found to preserve other key characteristics. For
instance, the degree of branching\cite{frey1999degree}, which is defined as:
\begin{equation}\label{eq:degreeofbranch}
    \mathrm{DB=2D/(2D+L),}
\end{equation}
where D and L are number of dendritic monomers (branching points)
and linear monomers, is also preserved along with the length
polydispersity. This is illustrated in both Fig.\ \ref{fig:db_distr}
and Table \ref{tab:polymerens}, which demonstrate that the
characteristics of both the topological and chain length
polydispersity are overall inherited from the large BA batch in the
sub-batches B$1$-B$4$. Note that, as explained in Section \ref{sec:mdmodel}, the macroinitiator and the $\mathrm{AB_{2}}$ monomers are considered solvophilic and compose the entire hyperbranched part of the polymer. The results are consistent with previous Monte-Carlo simulations\cite{Hanselmann1998}. 
\begin{table*}
\centering
\setlength{\tabcolsep}{0.2em} 
\renewcommand{\arraystretch}{1.5} 
\scriptsize
\begin{tabular}{|c|c|c|c|c|c|c|}
\hline
Target PDI & 1.0 & 1.1 & 1.2 & 1.3 & 1.4 & 1.5 \\ 
\hline
BA & 76.0(1.00)/0.56$\pm$0.05 & 76.1(1.10)/0.56$\pm$0.05 & 74.3(1.18)/0.56$\pm$0.05 & 75.6(1.33)/0.57$\pm$0.05 & 73.4(1.37)/0.57$\pm$0.05 & 74.7(1.56)/0.57$\pm$0.06 \\ 
\hline
B1 & 76.0(1.00)/0.56$\pm$0.05 & 76.1(1.10)/0.56$\pm$0.05 & 74.3(1.18)/0.56$\pm$0.05 & 75.6(1.34)/0.57$\pm$0.05 & 73.5(1.37)/0.57$\pm$0.05 & 74.7(1.56)/0.57$\pm$0.06 \\ 
\hline
B2 & 76.0(1.00)/0.56$\pm$0.04 & 76.2(1.10)/0.56$\pm$0.05 & 74.3(1.18)/0.56$\pm$0.05 & 75.6(1.33)/0.57$\pm$0.06 & 73.4(1.37)/0.56$\pm$0.05 & 74.7(1.56)/0.56$\pm$0.05 \\ 
\hline
B3 & 76.0(1.00)/0.56$\pm$0.04 & 76.2(1.10)/0.56$\pm$0.05 & 74.3(1.18)/0.56$\pm$0.05 & 75.6(1.33)/0.57$\pm$0.05 & 73.4(1.37)/0.57$\pm$0.05 & 74.7(1.55)/0.57$\pm$0.06 \\ 
\hline
B4 & 76.0(1.00)/0.56$\pm$0.05 & 76.0(1.10)/0.56$\pm$0.05 & 74.3(1.18)/0.56$\pm$0.05 & 75.6(1.32)/0.56$\pm$0.06 & 73.4(1.37)/0.56$\pm$0.06 & 74.7(1.55)/0.57$\pm$0.06 \\ 
\hline
\end{tabular}

\caption{Statistical properties of
polymers in each representative batch of LHBC molecules (see text),
with notation ''Average number of AB$_2$
monomers''(''PDI'')/''DB $\pm$
Error of DB''. Note that the calculation of DB only involves the solvophilic part of the polymer, while the calculation of the PDI involves only the $\mathrm{AB_{2}}$ monomers. Small deviations from the target values arise due to sampling.} 
\label{tab:polymerens}
\end{table*}
\subsection{Equilibrium micelle structures}\label{sec:eqmicresults}
To determine the equilibrium morphology and the CMC of the systems, we conducted one-,
two-, and three-dimensional SCF calculations in the grand canonical ensemble, which resulted in lamellae, cylindrical, or spherical micelles, respectively. For simplicity, we only consider solvophobic (H), solvophilic (P) and solvent (S) in these calculations and do not distinguish between the macroinitiator and the $\mathrm{AB_{2}}$ monomers, to which we all refer as P. In particular, we assume that all monomers have the same monomeric volume $v_P$.
We further assume that the polymers in
the micelle are in chemical equilibrium with a homogeneous solution (bath)
of chains of type $i$ with global average polymer volume fraction
$\Bar{\phi}$. The chemical potentials $\mu_i$ (Eq. (\ref{eq:chemicalpotential})) of each type, are then given in terms of ${w_{i}}$, which is the fraction of chains of type $i$ in the bath. 
We set ${w_{i}=1/n_{T}}$, meaning
that chains of all types $i$ are incorporated into the micelle with the same {\em a
priori} probability. The actual fraction of chains $i$ in the micelle
may of course differ from $w_i$.

\begin{figure}
    \centering
    \includegraphics[width=0.48\textwidth]{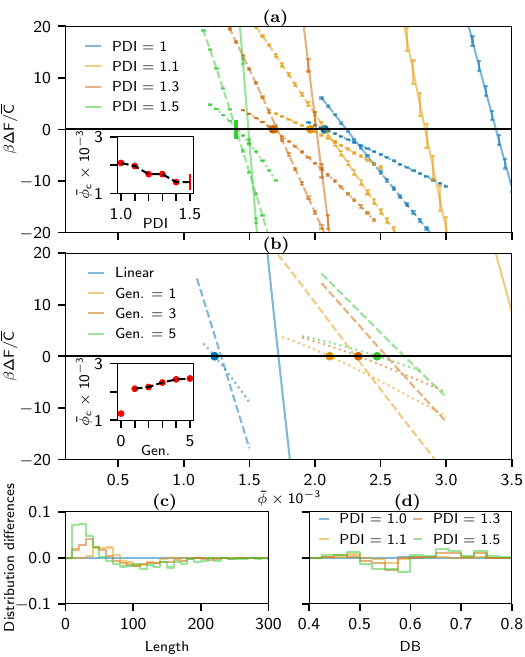}
    \caption{(a,b): Rescaled free energy difference $\mathrm{\Delta F}$ between homogeneous and
    inhomogeneous states against the average polymer volume fraction
    in the bath ${\Bar{\phi}}$ for polydisperse LHBCs (a)
    and monodisperse linear and LDBCs (b).
    Each color corresponds to a different polymer system, while the
    style of the line corresponds to lamella (solid line), cylindrical
    (dashed line) and spherical (dotted line) micelle states. For the
    LHBCs (a), errors obtained from averaging over batches are also shown. The insets in (a) and (b) show the state with the lowest critical
    concentration ${\Bar{\phi}_c}$ against PDI and number of
    generations, respectively. The circular (\protect\clcir) and
    rectangular (\protect\rect) symbols represent 
    spherical micelles and cylindrical micelles
    respectively as the equilibrium morphology, in both the main and inset plots. (c,d):
    Differences in the
    distribution of chain lengths and degree of branching between the
    bath and micelle were observed for LHBCs.
    The macromolecular weight and DB are calculated only for the hyperbranched
    part of the LHBCs.}
    \label{fig:cmcs} 
\end{figure}

First, we varied the average polymer volume fraction, ${\Bar{\phi}}$, of the bath and evaluated the
 free energy difference ($\mathrm{\Delta F}$)
between the inhomogeneous and homogeneous states for each case.
 Selected curves for $\mathrm{\Delta F}$ as a
function of $\Bar{\phi}$ are shown in Figs.\ \ref{fig:cmcs}a and
\ref{fig:cmcs}b. The critical volume fraction, $\Bar{\phi}_c$, is defined as the
lowest value of ${\bar{\phi}}$ among the three morphologies for
which $\mathrm{\Delta F} = 0$. This represents the lowest polymer
volume fraction at which micelles begin to form, with the
corresponding morphology being the equilibrium micelle morphology.
 The resulting values of $\Bar{\phi}_c$
and the respective morphologies are shown in the insets of Fig.\ \ref{fig:cmcs}: as a function of PDI for LHBCs
in Fig.\ \ref{fig:cmcs}a, and as a function of generations for
LDBCs in Fig.\ \ref{fig:cmcs}b. In
LDBC systems, $\Bar{\phi}_c$ increases with increasing generations,
consistent with prior findings\cite{Lebedeva2018}. In LHBC systems,
$\Bar{\phi}_c$ decreases with increasing PDI, which aligns with
observations from micelles formed by linear block copolymers with
a polydisperse solvophilic block\cite{Mantha2019}. For all polymer
systems tested, the equilibrium morphology was
spherical micelles, except for LHBCs at PDI = $1.5$,
which transitioned to cylindrical
micelles.
In Fig.\ \ref{fig:cmcs}c, the differences in the length distribution
of the solvophilic part between the micelle and the bath are shown for
the different LHBC systems. Greater polydispersity results in a larger proportion
of both smaller and larger chains in the bath. Smaller chains,
being overall more solvophobic, are preferentially attracted to the micelle, while larger
chains are preferred in the bath. Additionally, chains with
smaller solvophilic parts lose less configurational entropy upon
incorporation into micelles compared to larger chains, which explains the decrease in
${\bar{\phi}_c}$ with increasing PDI. The eventual
transition of the equilibrium morphology from spherical to
cylindrical can be attributed to 
smaller chains having a higher packing parameter\cite{israelachvili1976theory}.
Fig.\ \ref{fig:cmcs}d
demonstrates that the effects of topology are
minimal, as the differences in the degree of branching within the
micelle and the bath for the \mbox{PDI = $1$} case are negligible. At PDI = $1$, all chains have
equal molecular weight, so there is no size-based driving
force, unlike in the other cases.  Thus, the
differences observed for the other PDI cases can be primarily attributed to indirect effects of
molecular weight polydispersity, rather than topological
polydispersity.  However, this does not rule out a
potential impact of topological polydispersity in systems with fewer
chains than those tested.

\begin{figure}
    \centering
    \includegraphics[width=0.48\textwidth]{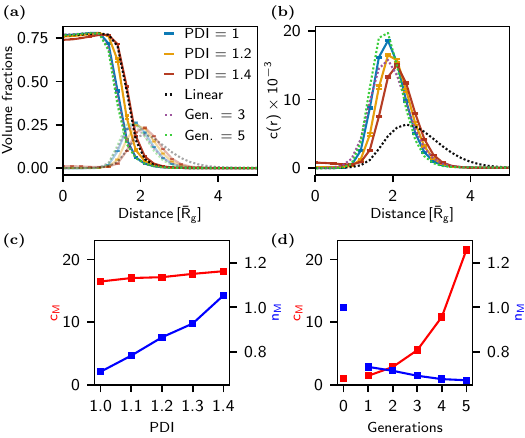}
     \caption{(a) Volume
     fraction profiles for spherical micelles at their CMC
     vs. the distance from the center. The
     opaque lines correspond to the solvophobic monomers while the
     translucent lines correspond to the solvophilic monomers. (b) Normalized average chain end
     profiles 
     vs. the distance from the center of the micelles. The colors and line style for each
     polymer system are shown in the legend. For the LHBC systems,
     errors are also shown. (c,d): Number of solvophilic chain ends in the micelle
     ${c_{M}}$ (red) and number of chains in the
     micelle ${n_{M}}$ (blue) normalized by the
     corresponding number for monodisperse linear
     diblock chains, for LHBCs as a function of PDI (c) and
     LDBCs as a function of generation number (d).}
     \label{fig:volfra_and_ends}
\end{figure}

Next, we compare the properties of the equilibrated spherical micelles.
We define the terminal end distribution 
${c(r)}$, the number of chains ${n_M}$, and the number of
terminal ends in the micelle, ${c_M}$, as:
\begin{align}\label{eq:chends}
  &c(\bb{r})=\sum_{i}^{n_{T}}\exp(\beta \mu_{i}
    +\ln(\Bar{N}))\\[3pt]
  &\times
     \sum_{\bi{1}{2}}^{TL}q_{\bi{1}{2}}
       (\bb{r},s^{\text{max}}_{\bi{1}{2}}) \:
      q^{\dagger}_{\bi{1}{2}}(\bb{r},s^{\text{max}}_{\bi{1}{2}})
        \nonumber\\[3pt]
  &c_{M}=\int_{\bb{V_{c.}}}\text{d}V\, c(\bb{r})/v_{P}\\[3pt]
  &n_{M}=\int_{\bb{V_{c.}}}\text{d}V\, \phi_{H}
    (\bb{r})/(N_{H}v_{P}),
\end{align}
where the sum is performed over the terminal blocks (TL) of chain type $i$,
${\bb{V_{c.}}}$ is a sphere with a cutoff radius of $6.0$
[${\Bar{R}_{g}}$] and ${N_{H}}$ is the length of the
solvophobic block. The solvophobic volume fraction $\phi_{H}$, the backward propagator $q$ and $q^{\dagger}$, as well as the notation are defined in Section \ref{sec:scfmodel}.
In Fig.\ \ref{fig:volfra_and_ends}a, the volume
fraction profiles of equilibrium micelles are shown for a
selection of polymer systems.  These profiles are only
marginally influenced by polydispersity, with the LHBC systems
exhibiting profiles that lie between those of the linear and
the other LDBC-based micelle systems. The results indicate that
the impact of polydispersity on the equilibrium volume fraction profiles
is relatively minor compared to other factors, such as polymer
architecture or the generation of LDBCs.  These findings are
consistent with a previous study on micelles formed by linear block
copolymers\cite{Mantha2019}, which showed that polydispersity in the
solvophobic block had a strong effect on the volume fraction profiles,
whereas polydispersity in the solvophilic block had little to no
effect. We note that the solvent content
in the micelle core is relatively high in these calculations, 
around $25$\%.  This is a consequence of the monomer-solvent interactions 
($\chi_{HS}=0.61$) being relatively small. This value is inspired 
by an empirical estimate for polystyrene in n-decane based 
on Hansen's solubility parameters\cite{Hansen_solubility}. The 
$\chi_{HS}$ values for hydrophobic components of pharmaceutical 
micelles in water, such as polylactate, are typically about twice as high,
therefore the solvent content in the micelle core will be lower.
However, this should not change the general trends.

An interesting effect is observed in the
high PDI case (PDI = $1.4$). As shown in Fig.\ \ref{fig:volfra_and_ends}, a sizable amount of solvophilic
monomers enters the predominantly solvophobic
core, leading to a corresponding decrease in the solvophobic
contribution. High PDIs can result in polymers
with small solvophilic contributions, which makes them nearly entirely
solvophobic. Consequently, the polymers tend to position themselves deeper within the micelle core.
As a result, polymers with small solvophilic blocks may occasionally
flip, with their solvophilic segments pointing inward rather than
outward. This flipping behavior explains the
reduced solvophobicity and the slight increase of the solvophilic
contributions within the core. 
It highlights the complex nature of micelle formation at high polydispersity, where the distribution of solvophobic
and solvophilic monomers becomes less predictable. 
The phenomenon may also explain the paradoxical
observation that the equilibrium micelle size increases with
increasing PDI, despite being composed of smaller chains. In other words, the shorter chains, which behave almost
entirely as solvophobic molecules, contribute to swelling of the
micelle core.  
When comparing the volume fraction profiles of
LHBC and LDBC micelles to those of micelles composed of linear chains,
as shown in Fig.\ \ref{fig:volfra_and_ends}a, one finds that the
LHBC micelles at high PDI exhibit the highest resemblance. In
contrast, the normalized terminal end
distributions in Fig.\ \ref{fig:volfra_and_ends}b present a
different picture. Here, LHBCs show
greater similarity to LDBCs than to their linear counterparts,
as they feature a much more concentrated corona. Increasing the PDI shifts the peak of the
distribution toward larger values, which is consistent with the
expected increase in micelle size.
Unexpectedly, the number of terminal ends, ${c_{M}}$, for LHBCs appears to remain
relatively constant with respect to PDI, showing only a minute
increase as PDI increases, as illustrated in
Fig.\ \ref{fig:volfra_and_ends}c (red curve). This results from the interplay of
two opposing factors: As can be seen in
Fig.\ \ref{fig:volfra_and_ends}c (blue curve), the number of
chains, ${n_{M}}$, increases with increasing PDI in LHBC micelles,
consistent with the increase in the micelle size discussed above. On the other hand, the ratio ${c_{M}/n_{M}}$,
which corresponds to the average number of terminal ends
per polymer in the micelle, decreases with PDI due to the preference for shorter chains
in such micelles. Therefore, despite larger PDIs leading to larger
micelles, which would typically result in a higher ${c_{M}}$,
the presence of shorter chains with fewer terminal ends keeps ${c_{M}}$ relatively unchanged. In LDBC
micelles, a similar competition arises. The number of terminal ends
per chain increases exponentially with increasing number of
generations, but the number of chains $n_M$ decreases (see Fig.\
\ref{fig:volfra_and_ends}d, blue curve). However, in this case, the
first effect dominates by far, such that the number of chain ends in
LDBC micelles still increases exponentially as a function of
the generation number (Fig.\ \ref{fig:volfra_and_ends}d, red
curve), at least up to the fifth generation.

\subsection{Micelle size and shape fluctuations}

After we discussed the properties of equilibrium micelles, we
now turn to the free energy penalties associated with deviations from
the preferred micelle size and shape. This analysis gives information
on the stability and polydispersity of micelles, and on their
resistance to deformations. The small
statistical errors observed in Figs.\ \ref{fig:cmcs} and
\ref{fig:volfra_and_ends} for LHBCs
indicate that a single sub-batch is sufficient to capture the
behavior of the entire ensemble. Therefore, from this point onward,
the results for the LHBC ensembles will be based on the B$1$ batch for
each PDI.

We first examine the energy difference ${F_{M}(R_{M})}$ between the
equilibrium spherical micelle and a micelle of radius ${R_{M}}$
in a bath with an average polymer volume fraction
${\Bar{\phi}_c}$. To this end, we introduce
a constraint potential in
Eq.\ \eqref{eq:conpot},
\begin{align}\label{eq:conpot}
    V_{\text{con}}[\phi_{A}]&
      =\frac{\kappa_{\text{con}}}{2v^{*}}
        (\int_{\bb{V_{\text{ell.}}}} 
        \text{d}\bb{r} \phi_{H}
        (\bb{r})-\phi_{\text{con}})^{2},
\end{align}
where the integral is performed over an ellipsoid of volume
${\bb{V_{\text{ell.}}}}$ concentered with the micelles. We note that the additional energy term from Eq.\ \eqref{eq:conpot} is not explicitly added to the free energy ${F_{M}(R_{M})}$, only the field contribution of this potential is included, as was similarly done in Mantha {\it et al}\cite{Mantha2019}. We define the micelle radius ($R_M$) as the radius at which
${\phi_{H}}=0.5$, and the radius is calculated post
hoc following the SCF calculations. 

\begin{figure}
    \centering
    \includegraphics[width=0.48\textwidth]{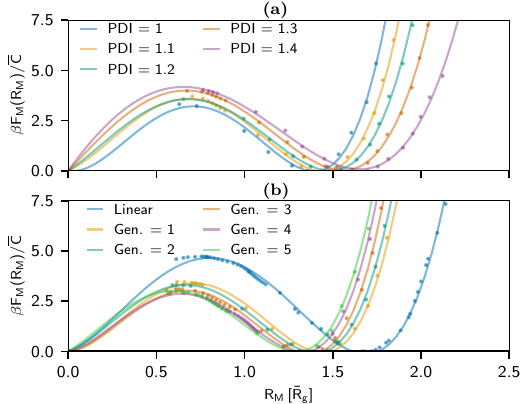}
    \caption{ Rescaled free energy of the micelle
    ${F_{M}(R_{M})}$ vs. the radius of the micelle ${R_{M}}$, for
    different PDIs (a) or linear and LDBCs (b) at their respective
    critical volume fractions ${\Bar{\phi}_c}$. The results were
    obtained by varying $\phi_{\text{con}}$ in the
    constraint potential of Eq.\ \eqref{eq:conpot}, while
    ${\kappa_{\text{con}}\bar{N}}=1$ and ${\bb{V_{\text{ell.}}}}$ which is a
    sphere of radius ${R/\Bar{R}_{g}=5}$, were kept 
    constant. A fourth order polynomial was fitted to each system.  
    \label{fig:size_dis}}
\end{figure}

The results, shown in Fig.\ \ref{fig:size_dis}, indicate that the most
stable micelles, as characterized by the height of the energy barrier,
are those composed of linear polymers, followed by the system with PDI = $1.4$. Increasing the PDI leads to a moderate increase in the energy
barrier, while for LDBCs, increasing the number of generations
slightly reduces it. Both behaviors can be attributed to the growing
and decreasing number of chains within the micelle for
increasing PDI and number of generations respectively.

Furthermore, we can inspect the curvature of $F_M(R_M)$ at the
minimum, which is related to the size distribution of micelles {\it
via} $P(R_M) \propto \exp(\beta F(R_M))$. For micelles composed of
monodisperse copolymers, the curvature appears to be largely independent
of the copolymer architecture. It is very similar for linear
copolymers, LDBCs, and LHBCs with PDI = $1$. However, if one increases the
PDI in the LHBC systems, the curvature decreases, indicating a
broadening of the micelle size distribution. We attribute this to the
greater number of smaller chains within the micelles. These smaller
chains contribute to the swelling of the micelles and help stabilize a
broader range of micelle sizes.

In a similar manner, we investigate the penalty associated
with deforming  the equilibrium
micelles, thus making them aspherical. To define this quantity
 we consider the
normalized moment of inertia tensor, which we define as
\begin{equation}\label{eq:inertialtens}
    I_{ij} = 
    \frac{\sum_p\phi_{H}^{(p)}
      (\lvert r_{(p)}\lvert^{2} \delta_{ij}
        -x_{i}^{(p)}x_{j}^{(p)})}
     {\sum_p\phi_{H}^{(p)}} { \,\text{for}\, i,j=1,\, 2,\, 3},
\end{equation}
where the sum over $p$ 
runs over all grid points obeying ${\phi_{H}}\geq0.05$, $x^{(p)}_{i}$ and $\lvert r_{(p)}\lvert$ are the Cartesian components and the distance from the center of the micelle respectively.
The asphericity of a micelle is then defined as:
\begin{equation}\label{eq:asphericity}
    A=\lambda_{z}^{2}-\frac{(\lambda_{x}^{2}+\lambda_{y}^{2})}{2},
\end{equation}
where $\lambda_{x,y,z}$ are the eigenvalues of the tensor in 
Eq.\ \eqref{eq:inertialtens}.

To impose different asphericities, we again use
the constraint potential defined in Eq.\ \eqref{eq:conpot}. However,
since this time, we wish to study the response of a given
equilibrium micelle to mechanical deformation, we fix the total number
of chains in the system, $n_b$, as well as the chain composition, and
perform the SCF calculations in the canonical ensemble.   
\begin{figure}
    \centering
    \includegraphics[width=0.48\textwidth]{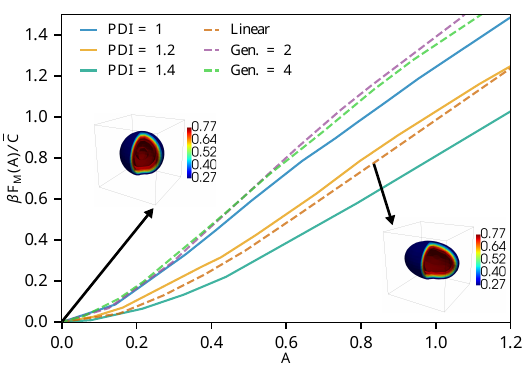}
    \caption{Free energy penalty $F_M$ for deforming a micelle from spherical to aspherical
    vs. asphericity $A$ for LHBC micelles (solid lines)
    and micelles composed of monodisperse LDBCs and linear
    copolymers  (dashed lines). To 
    enforce different asphericities, the constraint potential of
    Eq.\ \eqref{eq:conpot} was used, with 
    ${V_{\text{ell.}}}$  representing a spheroid
    with increasing size in the x-direction and decreasing size in the
    y and z directions such that ${V_{\text{ell.}}}$ matches the
    volume of the equilibrium micelle core, which was defined as the volume that obeys $\phi_{H}\ge0.5$. ${\kappa_{\text{con}}\bar{N}}$
    was set to $1$ while ${\phi_{\text{con}}}$ was determined by the
    volume contribution of the solvophobic micelle core at $A=0$.
    Also shown with corresponding arrows, are
    contour plots of the resulting morphology for the linear diblock
    results at $A=0$ and $A=0.84$.
    \label{fig:asphericities}
    }
\end{figure}
Fig.\ \ref{fig:asphericities} presents the energy
penalty for deforming micelles as a function of asphericity in the systems of interest along
with two example morphologies depicting the change from a spherical
micelle to a cigar like micelle. The figures
shows that LDBC  micelles and LHBC micelles at
PDI = $1$ case exhibit similar resistance to
deformation from their spherical shape,
 featuring the highest structural stability compared
to other, more deformable systems. This is expected, as the topology
of LDBCs aligns naturally with spherical micelles, and PDI = $1$
polymers, although slightly more flexible, mimic LDBCs. This minor increase in malleability in LHBC micelles at PDI = $1$ can be attributed to the diversity of polymer
topologies within the micelle, that can arrange themselves in
favorable positions so that for a given asphericity a smaller energy
penalty is paid. Conversely, increasing the PDI in LHBC
systems enhances the structural flexibility of the micelles,
making the PDI = $1.4$ system even more flexible than the linear micelle.

As the PDI increases, the diversity in chain topology and size also grows. Consequently, chains of different lengths adopt different spatial conformations within the micelle, as shown in Fig. \ref{fig:densxy}a,b. The increased diversity enhances adaptability to stress, since chains can adjust to micelle deformations by repositioning and reorientating. This is illustrated in Fig. \ref{fig:densxy}d which demonstrates that in a stretched micelle, the proportion of long chains oriented along the stretched axis, relative to short chains, is higher along the long axis than along the short axis.

\subsection{Encapsulation of solvophobic drug molecules}

\begin{figure}
    \centering
    \includegraphics[width=0.48\textwidth]{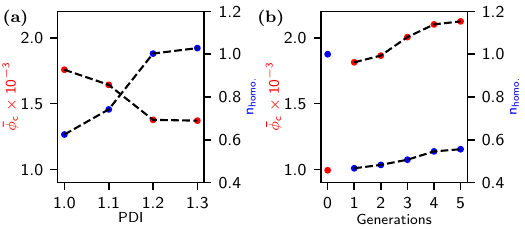}
    \caption{Critical volume fraction
    $\Bar{\phi}_{c}$ (excluding the contribution from the homopolymers
    $\Bar{\phi}_{h}$) (red) and average number of homopolymer chains 
 in the simulation box $n_{homo.}$, normalized by the corresponding number for monodisperse linear diblock chain (blue) vs. the polydispersity
    of LHBCs (a) and the number of generations of LDBCs (b). The homopolymer volume fraction in the reservoir solution is kept fixed at 
    $\Bar{\phi}_h = 10^{-5}$.
    \label{fig:encaps} }
\end{figure}

To conclude our investigation, we analyzed the encapsulation
properties of the micelle systems using a solvophobic homopolymer made
of type A monomers, with a length equivalent to the linear solvophobic
segment of each system. We fixed this homopolymer's contribution to
the total polymer volume fraction at ${\Bar{\phi}_h=10^{-5}}$, to
maintain consistent encapsulation conditions across all micelle
configurations. Simulations were then rerun for each system of
interest. The resulting data are presented
in Fig.\ \ref{fig:encaps}.

The CMC follows trends similar to those observed without
encapsulation; however, CMC values across all systems are lower, a
phenomenon commonly reported with the addition of solvophobic drugs \cite{zhang1996development}
Encapsulation also affects the equilibrium morphology in
LHBCs such that they adopt a cylindrical morphology already at \mbox{PDI = $1.4$}.

All LHBC systems exhibit superior encapsulation capacities compared with
LDBC systems, and increasing PDI further increases this capacity. We attribute
this to the fact that, as the homopolymer is incorporated within the micelle, the micelle swells and its size deviates from the equilibrium values in the absence of the drug, as illustrated in Fig. \ref{fig:radiiencaps}a. 
As we noted before, swelling is penalized more strongly in
LDBCs  compared with LHBCs (Fig.\ \ref{fig:size_dis}).
Indeed, Fig. \ref{fig:radiiencaps} shows that the micelle core size of LDBCs does not change upon incorporation of the homopolymer.
Therefore, LHBCs
can accommodate a higher payload due to their flexibility in size
fluctuations. 
\section{Conclusion}
We investigated
self-assembled micelles composed of polymers with a
monodisperse linear solvophobic block and a
solvophilic block of equal average molecular weight, which are either polydisperse hyperbranched (LHBC), monodisperse
dendritic (LDBC) or monodisperse linear (diblock). To do so, we first constructed a set of
polydisperse topologies for the hyperbranched case, mimicking the slow-monomer addition synthesis
protocol in MD simulations. Subsequently, we
continued our investigation using self-consistent field (SCF) numerical calculations. For this purpose, we developed a methodology that incorporates the
random branching characteristics of LHBCs, and we simulated these
systems in the grand canonical ensemble to account for the exchange
of polymers between micelles and their environment. 

We found that increasing the polydispersity in LHBCs improves
the stability of micelles and lowers the  critical
micelle concentration (CMC). This
effect is largely driven by smaller chains, that are relatively more
solvophobic, and therefore exhibit an increased
tendency to be incorporated into micelles.
In contrast, the topology of the polymers  appears to have a smaller impact in these systems, resulting in only slight
differences in the aforementioned characteristics compared to LDBC systems. 

Volume fraction profiles and terminal end distributions were also
found to be broadly similar between LHBC and LDBC micelles. However, the number of chain ends in
LHBC micelles was found to be surprisingly independent of
polydispersity, and comparable to micelles composed of LDBCs with five generations. This independence is attributed to a combination of two factors: an increase
in the number of chains within the micelle with increasing
polydispersity, and a simultaneous decrease in the average number
of terminal ends per chain in the
micelle due to the higher content of shorter
chains, which naturally have a smaller number of terminal ends. 

Our calculations suggest that
LHBC micelles are generally more diverse in size and offer
less resistance to deformations from their spherical shape compared to that of LDBC micelles. Finally, we probed the 
capability of the micelles to encapsulate
solvophobic drugs by testing them with a solvophobic homopolymer. We found
that, due to their increased malleability, LHBCs can accommodate
a larger payload than LDBCs, although encapsulation can influence
their equilibrium morphologies and may induce a transition from spherical to cylindrical morphologies.

In summary, we have demonstrated that LHBC micelles exhibit
behaviors similar to those of LDBC micelles,
with findings indicating that the random topology of LHBCs is not the
primary determinant of their characteristics.
The polydispersity in size plays a more significant role.
Additionally, the increased diversity in LHBCs  proves
advantageous, contributing to the enhanced encapsulation capacity and
improved stability. We believe, therefore, that the randomness inherent in LHBCs can be thought of not as a drawback, but as an attribute that can be explored and taken advantage of.

Future research could explore reverse micelles,
where the branched blocks form the core, as in this type of systems,
the influence of topology is expected to be more pronounced compared to the systems examined here\cite{Zhu2011,Tan2015}.
Additionally, exploring the effects of
terminal group modifications on these polymers could further refine
our understanding of micelle behavior. 
We have made our code available as part of the
SCF package published in Qiao {\it et al}
\cite{qiao}, which can be used to simulate multiblock copolymers of
any tree-like graph topology and is
parallelized for polydisperse systems.

\section{Data}
The data and code for the MD simulations and SCF calculations can be found at:
\url{https://gitlab.rlp.net/mgiannak/hyperbranched}.

\section*{Supporting Information}
Contains information about chain size distribution in micelles for different sizes (S1), volume fraction profiles for aspherical micelles (S2), change of micelle radius comparison with and without encapsulation of homopolymers (S3), free energy differences against free chain concentration with homopolymers encapsulation (S4) and correlation between order of time of addition and distance against the generation number (S5).\newline
\section*{Acknowledgments} 
This work was funded by the German Science
Foundation (DFG) within Grant number $446008821$, and the Agence
Nationale de La Recherche, France. Partial funding was also received
by the DFG within Grant number $429613790$.  M.G. is associate member of
the integrated graduate school of the collaborative research center
TRR 146 ''Multiscale modeling of soft matter systems'', grant number
$233630050$.

\begin{appendix}

\section{Additional calculations}\label{addcalc}
In homogeneous systems, the fields and propagators do not
vary spatially and the propagator equations, Eqs.\
\eqref{eq:mde},  can be solved analytically.
This results in:
\begin{equation*}
    q_{\bi{1}{2}}(s) q^{\dagger}_{\bi{1}{2}}(s)
    \equiv \frac{\Bar{Q_{i}}}{V} \quad \text{for all $\bi{1}{2}$ and
    $s$},
\end{equation*}\vspace{0.25cm}

and the following expression for the contribution of chain 
$i$ to the volume fraction of type $\alpha$:
\begin{eqnarray*}
\Bar{\phi}_{\alpha,i}&= v_{P}&
 \exp(\beta\mu_{i} + \ln(\Bar{N}))  \times
 \\ &&
 \sum_{\bi{1}{2}}
    \int \text{d}s \, 
   q_{\bi{1}{2}}(s)q^{\dagger}_{\bi{1}{2}}(s)
   \theta_{\alpha,\bi{1}{2}}
   \\
&=&v_{P}\exp(\beta\mu_{i}+\ln(\Bar{N}))
  \frac{\Bar{Q_{i}}}{V} f_{\alpha,i}\frac{N_{i}}{\Bar{N}},
\end{eqnarray*}
where $\theta_{\alpha,\bi{1}{2}}$ is oneaccurately if the block
$\bi{1}{2}$ has the type $\alpha$ 
and zero otherwise, the sum
$\bi{1}{2}$ runs over all blocks in the chain $i$ and 
$f_{\alpha,i}$ is the fraction of chain type $i$ that is of 
type $\alpha$.
The average volume contribution of the polymer $i$ of monomer type $\alpha$ can also be written as:
\begin{equation*}
    \Bar{\phi}_{\alpha,i}=\frac{w_{i}f_{\alpha,i}N_{i}}{\Bar{N}}\Bar{\phi}.
\end{equation*}
Equating the two equations 
above leads to Eq.\ \eqref{eq:chemicalpotential}.
We should note that Eq.\ \eqref{eq:chemicalpotential} remains
valid even if $\Bar{Q}_i$ is not evaluated in SCF approximation, but
by more sophisticated means. Taking into account effects of nonideal
chain conformations, {\it e.g.}, due to the fact that solvophobic
blocks of isolated chains might collapse\cite{Wang2012}, would shift
the values of $\Bar{Q}_i$ and hence $\mu_i$. Here, we neglect such
effects for consistency. In full inhomogeneous SCF calculations, the
micelles are also surrounded by a homogeneous solution, and we design
the study such that this solution is equivalent to the reservoir
solution.
\end{appendix}

\bibliography{bibliography}

\providecommand{\noopsort}[1]{}\providecommand{\singleletter}[1]{#1}%
\begin{thebibliography}{81}%
\makeatletter
\providecommand \@ifxundefined [1]{%
 \@ifx{#1\undefined}
}%
\providecommand \@ifnum [1]{%
 \ifnum #1\expandafter \@firstoftwo
 \else \expandafter \@secondoftwo
 \fi
}%
\providecommand \@ifx [1]{%
 \ifx #1\expandafter \@firstoftwo
 \else \expandafter \@secondoftwo
 \fi
}%
\providecommand \natexlab [1]{#1}%
\providecommand \enquote  [1]{``#1''}%
\providecommand \bibnamefont  [1]{#1}%
\providecommand \bibfnamefont [1]{#1}%
\providecommand \citenamefont [1]{#1}%
\providecommand \href@noop [0]{\@secondoftwo}%
\providecommand \href [0]{\begingroup \@sanitize@url \@href}%
\providecommand \@href[1]{\@@startlink{#1}\@@href}%
\providecommand \@@href[1]{\endgroup#1\@@endlink}%
\providecommand \@sanitize@url [0]{\catcode `\\12\catcode `\$12\catcode
  `\&12\catcode `\#12\catcode `\^12\catcode `\_12\catcode `\%12\relax}%
\providecommand \@@startlink[1]{}%
\providecommand \@@endlink[0]{}%
\providecommand \url  [0]{\begingroup\@sanitize@url \@url }%
\providecommand \@url [1]{\endgroup\@href {#1}{\urlprefix }}%
\providecommand \urlprefix  [0]{URL }%
\providecommand \Eprint [0]{\href }%
\providecommand \doibase [0]{http://dx.doi.org/}%
\providecommand \selectlanguage [0]{\@gobble}%
\providecommand \bibinfo  [0]{\@secondoftwo}%
\providecommand \bibfield  [0]{\@secondoftwo}%
\providecommand \translation [1]{[#1]}%
\providecommand \BibitemOpen [0]{}%
\providecommand \bibitemStop [0]{}%
\providecommand \bibitemNoStop [0]{.\EOS\space}%
\providecommand \EOS [0]{\spacefactor3000\relax}%
\providecommand \BibitemShut  [1]{\csname bibitem#1\endcsname}%
\let\auto@bib@innerbib\@empty
\bibitem [{\citenamefont {Hamley}(1998)}]{hamley1998physics}%
  \BibitemOpen
  \bibfield  {author} {\bibinfo {author} {\bibfnamefont {Ian~W}\ \bibnamefont
  {Hamley}},\ }\href@noop {} {\emph {\bibinfo {title} {The physics of block
  copolymers}}}\ (\bibinfo  {publisher} {Oxford University Press},\ \bibinfo
  {year} {1998})\BibitemShut {NoStop}%
\bibitem [{\citenamefont {Matsen}\ and\ \citenamefont
  {Schick}(1994)}]{Matsen1994}%
  \BibitemOpen
  \bibfield  {author} {\bibinfo {author} {\bibfnamefont {M.~W.}\ \bibnamefont
  {Matsen}}\ and\ \bibinfo {author} {\bibfnamefont {M.}~\bibnamefont
  {Schick}},\ }\bibfield  {title} {\enquote {\bibinfo {title} {{Microphases of
  a Diblock Copolymer with Conformational Asymmetry}},}\ }\href {\doibase
  10.1021/MA00092A049/ASSET/MA00092A049.FP.PNG_V03} {\bibfield  {journal}
  {\bibinfo  {journal} {Macromolecules}\ }\textbf {\bibinfo {volume} {27}},\
  \bibinfo {pages} {4014--4015} (\bibinfo {year} {1994})}\BibitemShut {NoStop}%
\bibitem [{\citenamefont {Khandpur}\ \emph {et~al.}(1995)\citenamefont
  {Khandpur}, \citenamefont {F\"orster}, \citenamefont {Bates}, \citenamefont
  {Hamley}, \citenamefont {Ryan}, \citenamefont {Almdal},\ and\ \citenamefont
  {Mortensen}}]{Khandpur1995}%
  \BibitemOpen
  \bibfield  {author} {\bibinfo {author} {\bibfnamefont {Ashish~K}\
  \bibnamefont {Khandpur}}, \bibinfo {author} {\bibfnamefont {Stephan}\
  \bibnamefont {F\"orster}}, \bibinfo {author} {\bibfnamefont {Frank~S}\
  \bibnamefont {Bates}}, \bibinfo {author} {\bibfnamefont {Ian~W}\ \bibnamefont
  {Hamley}}, \bibinfo {author} {\bibfnamefont {Anthony~J}\ \bibnamefont
  {Ryan}}, \bibinfo {author} {\bibfnamefont {Kristoffer}\ \bibnamefont
  {Almdal}}, \ and\ \bibinfo {author} {\bibfnamefont {Kell}\ \bibnamefont
  {Mortensen}},\ }\bibfield  {title} {\enquote {\bibinfo {title}
  {{Polyisoprene-Polystyrene Diblock Copolymer Phase Diagram near the
  Order-Disorder Transition}},}\ }\href
  {https://pubs.acs.org/sharingguidelines} {\bibfield  {journal} {\bibinfo
  {journal} {Macromolecules}\ }\textbf {\bibinfo {volume} {28}},\ \bibinfo
  {pages} {8796--8806} (\bibinfo {year} {1995})}\BibitemShut {NoStop}%
\bibitem [{\citenamefont {Mai}\ and\ \citenamefont
  {Eisenberg}(2012)}]{Mai2012}%
  \BibitemOpen
  \bibfield  {author} {\bibinfo {author} {\bibfnamefont {Yiyong}\ \bibnamefont
  {Mai}}\ and\ \bibinfo {author} {\bibfnamefont {Adi}\ \bibnamefont
  {Eisenberg}},\ }\bibfield  {title} {\enquote {\bibinfo {title}
  {{Self-assembly of block copolymers}},}\ }\href {\doibase 10.1039/C2CS35115C}
  {\bibfield  {journal} {\bibinfo  {journal} {Chemical Society Reviews}\
  }\textbf {\bibinfo {volume} {41}},\ \bibinfo {pages} {5969--5985} (\bibinfo
  {year} {2012})}\BibitemShut {NoStop}%
\bibitem [{\citenamefont {Bates}\ \emph {et~al.}(2019)\citenamefont {Bates},
  \citenamefont {Lequieu}, \citenamefont {Barbon}, \citenamefont {Lewis},
  \citenamefont {Delaney}, \citenamefont {Anastasaki}, \citenamefont {Hawker},
  \citenamefont {Fredrickson},\ and\ \citenamefont {Bates}}]{Bates2019}%
  \BibitemOpen
  \bibfield  {author} {\bibinfo {author} {\bibfnamefont {Morgan~W.}\
  \bibnamefont {Bates}}, \bibinfo {author} {\bibfnamefont {Joshua}\
  \bibnamefont {Lequieu}}, \bibinfo {author} {\bibfnamefont {Stephanie~M.}\
  \bibnamefont {Barbon}}, \bibinfo {author} {\bibfnamefont {Ronald~M.}\
  \bibnamefont {Lewis}}, \bibinfo {author} {\bibfnamefont {Kris~T.}\
  \bibnamefont {Delaney}}, \bibinfo {author} {\bibfnamefont {Athina}\
  \bibnamefont {Anastasaki}}, \bibinfo {author} {\bibfnamefont {Craig~J.}\
  \bibnamefont {Hawker}}, \bibinfo {author} {\bibfnamefont {Glenn~H.}\
  \bibnamefont {Fredrickson}}, \ and\ \bibinfo {author} {\bibfnamefont
  {Christopher~M.}\ \bibnamefont {Bates}},\ }\bibfield  {title} {\enquote
  {\bibinfo {title} {{Stability of the A15 phase in diblock copolymer
  melts}},}\ }\href {\doibase
  10.1073/PNAS.1900121116/SUPPL_FILE/PNAS.1900121116.SAPP.PDF} {\bibfield
  {journal} {\bibinfo  {journal} {Proceedings of the National Academy of
  Sciences of the United States of America}\ }\textbf {\bibinfo {volume}
  {116}},\ \bibinfo {pages} {13194--13199} (\bibinfo {year}
  {2019})}\BibitemShut {NoStop}%
\bibitem [{\citenamefont {Meier}(1969)}]{Meier1969}%
  \BibitemOpen
  \bibfield  {author} {\bibinfo {author} {\bibfnamefont {D.~J.}\ \bibnamefont
  {Meier}},\ }\bibfield  {title} {\enquote {\bibinfo {title} {{Theory of block
  copolymers. I. Domain formation in A-B block copolymers}},}\ }\href {\doibase
  10.1002/POLC.5070260106} {\bibfield  {journal} {\bibinfo  {journal} {Journal
  of Polymer Science Part C: Polymer Symposia}\ }\textbf {\bibinfo {volume}
  {26}},\ \bibinfo {pages} {81--98} (\bibinfo {year} {1969})}\BibitemShut
  {NoStop}%
\bibitem [{\citenamefont {{Anal Chim Acta}}\ \emph {et~al.}(1986)\citenamefont
  {{Anal Chim Acta}}, \citenamefont {in~Press}, \citenamefont {Ohta},\ and\
  \citenamefont {Kawasaki}}]{AnalChimActa1986}%
  \BibitemOpen
  \bibfield  {author} {\bibinfo {author} {\bibfnamefont {J~C}\ \bibnamefont
  {{Anal Chim Acta}}}, \bibinfo {author} {\bibnamefont {in~Press}}, \bibinfo
  {author} {\bibfnamefont {Takao}\ \bibnamefont {Ohta}}, \ and\ \bibinfo
  {author} {\bibfnamefont {Kyozi}\ \bibnamefont {Kawasaki}},\ }\bibfield
  {title} {\enquote {\bibinfo {title} {{Equilibrium Morphology of Block
  Copolymer Melts}},}\ }\href {\doibase 10.1021/MA00164A028} {\bibfield
  {journal} {\bibinfo  {journal} {Macromolecules}\ }\textbf {\bibinfo {volume}
  {19}},\ \bibinfo {pages} {2621--2632} (\bibinfo {year} {1986})}\BibitemShut
  {NoStop}%
\bibitem [{\citenamefont {Park}\ \emph {et~al.}(1997)\citenamefont {Park},
  \citenamefont {Harrison}, \citenamefont {Chaikin}, \citenamefont {Register},\
  and\ \citenamefont {Adamson}}]{Park1997}%
  \BibitemOpen
  \bibfield  {author} {\bibinfo {author} {\bibfnamefont {Miri}\ \bibnamefont
  {Park}}, \bibinfo {author} {\bibfnamefont {Christopher}\ \bibnamefont
  {Harrison}}, \bibinfo {author} {\bibfnamefont {Paul~M.}\ \bibnamefont
  {Chaikin}}, \bibinfo {author} {\bibfnamefont {Richard~A.}\ \bibnamefont
  {Register}}, \ and\ \bibinfo {author} {\bibfnamefont {Douglas~H.}\
  \bibnamefont {Adamson}},\ }\bibfield  {title} {\enquote {\bibinfo {title}
  {{Block copolymer lithography: Periodic arrays of $\sim$1011 holes in 1
  square centimeter}},}\ }\href {\doibase 10.1126/SCIENCE.276.5317.1401}
  {\bibfield  {journal} {\bibinfo  {journal} {Science}\ }\textbf {\bibinfo
  {volume} {276}},\ \bibinfo {pages} {1401--1404} (\bibinfo {year}
  {1997})}\BibitemShut {NoStop}%
\bibitem [{\citenamefont {Thurn-Albrecht}\ \emph {et~al.}(2000)\citenamefont
  {Thurn-Albrecht}, \citenamefont {Derouchey}, \citenamefont {Russell},\ and\
  \citenamefont {Jaeger}}]{Thurn-Albrecht2000}%
  \BibitemOpen
  \bibfield  {author} {\bibinfo {author} {\bibfnamefont {T.}~\bibnamefont
  {Thurn-Albrecht}}, \bibinfo {author} {\bibfnamefont {J.}~\bibnamefont
  {Derouchey}}, \bibinfo {author} {\bibfnamefont {T.~P.}\ \bibnamefont
  {Russell}}, \ and\ \bibinfo {author} {\bibfnamefont {H.~M.}\ \bibnamefont
  {Jaeger}},\ }\bibfield  {title} {\enquote {\bibinfo {title} {{Overcoming
  interfacial interactions with electric fields}},}\ }\href {\doibase
  10.1021/MA991896Z/ASSET/IMAGES/LARGE/MA991896ZF00005.JPEG} {\bibfield
  {journal} {\bibinfo  {journal} {Macromolecules}\ }\textbf {\bibinfo {volume}
  {33}},\ \bibinfo {pages} {3250--3253} (\bibinfo {year} {2000})}\BibitemShut
  {NoStop}%
\bibitem [{\citenamefont {Krausch}\ and\ \citenamefont
  {Magerle}(2002)}]{Krausch2002}%
  \BibitemOpen
  \bibfield  {author} {\bibinfo {author} {\bibfnamefont {Georg}\ \bibnamefont
  {Krausch}}\ and\ \bibinfo {author} {\bibfnamefont {Robert}\ \bibnamefont
  {Magerle}},\ }\bibfield  {title} {\enquote {\bibinfo {title} {{Nanostructured
  thin films via self-assembly of block copolymers}},}\ }\href {\doibase
  10.1002/1521-4095(20021104)14:21<1579::AID-ADMA1579>3.0.CO;2-6} {\bibfield
  {journal} {\bibinfo  {journal} {Advanced Materials}\ }\textbf {\bibinfo
  {volume} {14}},\ \bibinfo {pages} {1579--1583} (\bibinfo {year}
  {2002})}\BibitemShut {NoStop}%
\bibitem [{\citenamefont {Ahn}\ \emph {et~al.}(2014)\citenamefont {Ahn},
  \citenamefont {Park}, \citenamefont {Kim}, \citenamefont {Yoo}, \citenamefont
  {Ryu},\ and\ \citenamefont {Russell}}]{Ahn2014}%
  \BibitemOpen
  \bibfield  {author} {\bibinfo {author} {\bibfnamefont {Hyungju}\ \bibnamefont
  {Ahn}}, \bibinfo {author} {\bibfnamefont {Sungmin}\ \bibnamefont {Park}},
  \bibinfo {author} {\bibfnamefont {Sang~Woo}\ \bibnamefont {Kim}}, \bibinfo
  {author} {\bibfnamefont {Pil~J.}\ \bibnamefont {Yoo}}, \bibinfo {author}
  {\bibfnamefont {Du~Yeol}\ \bibnamefont {Ryu}}, \ and\ \bibinfo {author}
  {\bibfnamefont {Thomas~P.}\ \bibnamefont {Russell}},\ }\bibfield  {title}
  {\enquote {\bibinfo {title} {{Nanoporous block copolymer membranes for
  ultrafiltration: A simple approach to size tunability}},}\ }\href {\doibase
  10.1021/NN505234V/SUPPL_FILE/NN505234V_SI_001.PDF} {\bibfield  {journal}
  {\bibinfo  {journal} {ACS Nano}\ }\textbf {\bibinfo {volume} {8}},\ \bibinfo
  {pages} {11745--11752} (\bibinfo {year} {2014})}\BibitemShut {NoStop}%
\bibitem [{\citenamefont {Lazzari}\ and\ \citenamefont {{Arturo
  L{\'{o}}pez-Quintela}}(2003)}]{Lazzari2003}%
  \BibitemOpen
  \bibfield  {author} {\bibinfo {author} {\bibfnamefont {Massimo}\ \bibnamefont
  {Lazzari}}\ and\ \bibinfo {author} {\bibfnamefont {M.}~\bibnamefont {{Arturo
  L{\'{o}}pez-Quintela}}},\ }\bibfield  {title} {\enquote {\bibinfo {title}
  {{Block Copolymers as a Tool for Nanomaterial Fabrication}},}\ }\href
  {\doibase 10.1002/ADMA.200300382} {\bibfield  {journal} {\bibinfo  {journal}
  {Advanced Materials}\ }\textbf {\bibinfo {volume} {15}},\ \bibinfo {pages}
  {1583--1594} (\bibinfo {year} {2003})}\BibitemShut {NoStop}%
\bibitem [{\citenamefont {Kim}\ \emph {et~al.}(2010)\citenamefont {Kim},
  \citenamefont {Yang}, \citenamefont {Lee},\ and\ \citenamefont
  {Kim}}]{Kim2010}%
  \BibitemOpen
  \bibfield  {author} {\bibinfo {author} {\bibfnamefont {Jin~Kon}\ \bibnamefont
  {Kim}}, \bibinfo {author} {\bibfnamefont {Seung~Yun}\ \bibnamefont {Yang}},
  \bibinfo {author} {\bibfnamefont {Youngmin}\ \bibnamefont {Lee}}, \ and\
  \bibinfo {author} {\bibfnamefont {Youngsuk}\ \bibnamefont {Kim}},\ }\bibfield
   {title} {\enquote {\bibinfo {title} {{Functional nanomaterials based on
  block copolymer self-assembly}},}\ }\href {\doibase
  10.1016/J.PROGPOLYMSCI.2010.06.002} {\bibfield  {journal} {\bibinfo
  {journal} {Progress in Polymer Science}\ }\textbf {\bibinfo {volume} {35}},\
  \bibinfo {pages} {1325--1349} (\bibinfo {year} {2010})}\BibitemShut {NoStop}%
\bibitem [{\citenamefont {Lodge}\ \emph {et~al.}(2002)\citenamefont {Lodge},
  \citenamefont {Pudil},\ and\ \citenamefont {Hanley}}]{Lodge2002}%
  \BibitemOpen
  \bibfield  {author} {\bibinfo {author} {\bibfnamefont {Timothy~P.}\
  \bibnamefont {Lodge}}, \bibinfo {author} {\bibfnamefont {Bryant}\
  \bibnamefont {Pudil}}, \ and\ \bibinfo {author} {\bibfnamefont {Kenneth~J.}\
  \bibnamefont {Hanley}},\ }\bibfield  {title} {\enquote {\bibinfo {title}
  {{The full phase behavior for block copolymers in solvents of varying
  selectivity}},}\ }\href {\doibase
  10.1021/MA0200975/ASSET/IMAGES/LARGE/MA0200975F00013.JPEG} {\bibfield
  {journal} {\bibinfo  {journal} {Macromolecules}\ }\textbf {\bibinfo {volume}
  {35}},\ \bibinfo {pages} {4707--4717} (\bibinfo {year} {2002})}\BibitemShut
  {NoStop}%
\bibitem [{\citenamefont {Hanley}\ \emph {et~al.}(2000)\citenamefont {Hanley},
  \citenamefont {Lodge},\ and\ \citenamefont {Huang}}]{Hanley2000}%
  \BibitemOpen
  \bibfield  {author} {\bibinfo {author} {\bibfnamefont {Kenneth~J.}\
  \bibnamefont {Hanley}}, \bibinfo {author} {\bibfnamefont {Timothy~P.}\
  \bibnamefont {Lodge}}, \ and\ \bibinfo {author} {\bibfnamefont {Ching~I.}\
  \bibnamefont {Huang}},\ }\bibfield  {title} {\enquote {\bibinfo {title}
  {{Phase behavior of a block copolymer in solvents of varying selectivity}},}\
  }\href {\doibase 10.1021/MA000318B/ASSET/IMAGES/LARGE/MA000318BF00015.JPEG}
  {\bibfield  {journal} {\bibinfo  {journal} {Macromolecules}\ }\textbf
  {\bibinfo {volume} {33}},\ \bibinfo {pages} {5918--5931} (\bibinfo {year}
  {2000})}\BibitemShut {NoStop}%
\bibitem [{\citenamefont {Battaglia}\ and\ \citenamefont
  {Ryan}(2006)}]{Battaglia2006}%
  \BibitemOpen
  \bibfield  {author} {\bibinfo {author} {\bibfnamefont {Giuseppe}\
  \bibnamefont {Battaglia}}\ and\ \bibinfo {author} {\bibfnamefont
  {Anthony~J.}\ \bibnamefont {Ryan}},\ }\bibfield  {title} {\enquote {\bibinfo
  {title} {{Effect of amphiphile size on the transformation from a lyotropic
  gel to a vesicular dispersion}},}\ }\href {\doibase
  10.1021/MA052108A/ASSET/IMAGES/LARGE/MA052108AF00010.JPEG} {\bibfield
  {journal} {\bibinfo  {journal} {Macromolecules}\ }\textbf {\bibinfo {volume}
  {39}},\ \bibinfo {pages} {798--805} (\bibinfo {year} {2006})}\BibitemShut
  {NoStop}%
\bibitem [{\citenamefont {Barnhill}\ \emph {et~al.}(2015)\citenamefont
  {Barnhill}, \citenamefont {Bell}, \citenamefont {Patterson}, \citenamefont
  {Olds},\ and\ \citenamefont {Gianneschi}}]{Barnhill2015}%
  \BibitemOpen
  \bibfield  {author} {\bibinfo {author} {\bibfnamefont {Sarah~A.}\
  \bibnamefont {Barnhill}}, \bibinfo {author} {\bibfnamefont {Nia~C.}\
  \bibnamefont {Bell}}, \bibinfo {author} {\bibfnamefont {Joseph~P.}\
  \bibnamefont {Patterson}}, \bibinfo {author} {\bibfnamefont {Daniel~P.}\
  \bibnamefont {Olds}}, \ and\ \bibinfo {author} {\bibfnamefont {Nathan~C.}\
  \bibnamefont {Gianneschi}},\ }\bibfield  {title} {\enquote {\bibinfo {title}
  {{Phase diagrams of polynorbornene amphiphilic block copolymers in
  solution}},}\ }\href {\doibase
  10.1021/MA502163J/SUPPL_FILE/MA502163J_SI_001.PDF} {\bibfield  {journal}
  {\bibinfo  {journal} {Macromolecules}\ }\textbf {\bibinfo {volume} {48}},\
  \bibinfo {pages} {1152--1161} (\bibinfo {year} {2015})}\BibitemShut {NoStop}%
\bibitem [{\citenamefont {Gadelle}\ \emph {et~al.}(1995)\citenamefont
  {Gadelle}, \citenamefont {Koros},\ and\ \citenamefont
  {Schechter}}]{Gadelle1995}%
  \BibitemOpen
  \bibfield  {author} {\bibinfo {author} {\bibfnamefont {Frederic}\
  \bibnamefont {Gadelle}}, \bibinfo {author} {\bibfnamefont {William~J}\
  \bibnamefont {Koros}}, \ and\ \bibinfo {author} {\bibfnamefont {Robert~S}\
  \bibnamefont {Schechter}},\ }\bibfield  {title} {\enquote {\bibinfo {title}
  {Solubilization of aromatic solutes in block copolymers},}\ }\href
  {https://pubs.acs.org/sharingguidelines} {\bibfield  {journal} {\bibinfo
  {journal} {Macromolecules}\ }\textbf {\bibinfo {volume} {28}},\ \bibinfo
  {pages} {4883--4892} (\bibinfo {year} {1995})}\BibitemShut {NoStop}%
\bibitem [{\citenamefont {Sakai}\ and\ \citenamefont
  {Alexandridis}(2004)}]{Sakai2004}%
  \BibitemOpen
  \bibfield  {author} {\bibinfo {author} {\bibfnamefont {Toshio}\ \bibnamefont
  {Sakai}}\ and\ \bibinfo {author} {\bibfnamefont {Paschalis}\ \bibnamefont
  {Alexandridis}},\ }\bibfield  {title} {\enquote {\bibinfo {title}
  {{Single-step synthesis and stabilization of metal nanoparticles in aqueous
  pluronic block copolymer solutions at ambient temperature}},}\ }\href
  {\doibase 10.1021/LA049514S/ASSET/IMAGES/LARGE/LA049514SF00004.JPEG}
  {\bibfield  {journal} {\bibinfo  {journal} {Langmuir}\ }\textbf {\bibinfo
  {volume} {20}},\ \bibinfo {pages} {8426--8430} (\bibinfo {year}
  {2004})}\BibitemShut {NoStop}%
\bibitem [{\citenamefont {Peters}\ \emph {et~al.}(2012)\citenamefont {Peters},
  \citenamefont {Louzao},\ and\ \citenamefont {{Van Hest}}}]{Peters2012}%
  \BibitemOpen
  \bibfield  {author} {\bibinfo {author} {\bibfnamefont {Ruud~J.R.W.}\
  \bibnamefont {Peters}}, \bibinfo {author} {\bibfnamefont {Iria}\ \bibnamefont
  {Louzao}}, \ and\ \bibinfo {author} {\bibfnamefont {Jan~C.M.}\ \bibnamefont
  {{Van Hest}}},\ }\bibfield  {title} {\enquote {\bibinfo {title} {{From
  polymeric nanoreactors to artificial organelles}},}\ }\href {\doibase
  10.1039/C2SC00803C} {\bibfield  {journal} {\bibinfo  {journal} {Chemical
  Science}\ }\textbf {\bibinfo {volume} {3}},\ \bibinfo {pages} {335--342}
  (\bibinfo {year} {2012})}\BibitemShut {NoStop}%
\bibitem [{\citenamefont {Kataoka}\ \emph {et~al.}(2012)\citenamefont
  {Kataoka}, \citenamefont {Harada},\ and\ \citenamefont
  {Nagasaki}}]{Kataoka2012}%
  \BibitemOpen
  \bibfield  {author} {\bibinfo {author} {\bibfnamefont {Kazunori}\
  \bibnamefont {Kataoka}}, \bibinfo {author} {\bibfnamefont {Atsushi}\
  \bibnamefont {Harada}}, \ and\ \bibinfo {author} {\bibfnamefont {Yukio}\
  \bibnamefont {Nagasaki}},\ }\bibfield  {title} {\enquote {\bibinfo {title}
  {{Block copolymer micelles for drug delivery: Design, characterization and
  biological significance}},}\ }\href {\doibase 10.1016/J.ADDR.2012.09.013}
  {\bibfield  {journal} {\bibinfo  {journal} {Advanced Drug Delivery Reviews}\
  }\textbf {\bibinfo {volume} {64}},\ \bibinfo {pages} {37--48} (\bibinfo
  {year} {2012})}\BibitemShut {NoStop}%
\bibitem [{\citenamefont {Bose}\ \emph {et~al.}(2021)\citenamefont {Bose},
  \citenamefont {Burman}, \citenamefont {Sikdar},\ and\ \citenamefont
  {Patra}}]{Bose2021}%
  \BibitemOpen
  \bibfield  {author} {\bibinfo {author} {\bibfnamefont {Anamika}\ \bibnamefont
  {Bose}}, \bibinfo {author} {\bibfnamefont {Debanwita~Roy}\ \bibnamefont
  {Burman}}, \bibinfo {author} {\bibfnamefont {Bismayan}\ \bibnamefont
  {Sikdar}}, \ and\ \bibinfo {author} {\bibfnamefont {Prasun}\ \bibnamefont
  {Patra}},\ }\bibfield  {title} {\enquote {\bibinfo {title} {{Nanomicelles:
  Types, properties and applications in drug delivery}},}\ }\href {\doibase
  10.1049/NBT2.12018} {\bibfield  {journal} {\bibinfo  {journal} {IET
  Nanobiotechnology}\ }\textbf {\bibinfo {volume} {15}},\ \bibinfo {pages}
  {19--27} (\bibinfo {year} {2021})}\BibitemShut {NoStop}%
\bibitem [{\citenamefont {Hamley}(2005)}]{Hamley2005}%
  \BibitemOpen
  \bibfield  {author} {\bibinfo {author} {\bibfnamefont {Ian~W}\ \bibnamefont
  {Hamley}},\ }\href@noop {} {\emph {\bibinfo {title} {{Block copolymers in
  solution: fundamentals and applications}}}},\ \bibinfo {number} {112}\
  (\bibinfo  {publisher} {John Wiley \& Sons},\ \bibinfo {year}
  {2005})\BibitemShut {NoStop}%
\bibitem [{\citenamefont {Owen}\ \emph {et~al.}(2012)\citenamefont {Owen},
  \citenamefont {Chan},\ and\ \citenamefont {Shoichet}}]{Owen2012}%
  \BibitemOpen
  \bibfield  {author} {\bibinfo {author} {\bibfnamefont {Shawn~C.}\
  \bibnamefont {Owen}}, \bibinfo {author} {\bibfnamefont {Dianna~P.Y.}\
  \bibnamefont {Chan}}, \ and\ \bibinfo {author} {\bibfnamefont {Molly~S.}\
  \bibnamefont {Shoichet}},\ }\bibfield  {title} {\enquote {\bibinfo {title}
  {{Polymeric micelle stability}},}\ }\href {\doibase
  10.1016/J.NANTOD.2012.01.002} {\bibfield  {journal} {\bibinfo  {journal}
  {Nano Today}\ }\textbf {\bibinfo {volume} {7}},\ \bibinfo {pages} {53--65}
  (\bibinfo {year} {2012})}\BibitemShut {NoStop}%
\bibitem [{\citenamefont {Kelley}\ \emph {et~al.}(2013)\citenamefont {Kelley},
  \citenamefont {Albert}, \citenamefont {Sullivan},\ and\ \citenamefont
  {Epps}}]{Kelley2013}%
  \BibitemOpen
  \bibfield  {author} {\bibinfo {author} {\bibfnamefont {Elizabeth~G.}\
  \bibnamefont {Kelley}}, \bibinfo {author} {\bibfnamefont {Julie~N.L.}\
  \bibnamefont {Albert}}, \bibinfo {author} {\bibfnamefont {Millicent~O.}\
  \bibnamefont {Sullivan}}, \ and\ \bibinfo {author} {\bibfnamefont
  {Thomas~H.}\ \bibnamefont {Epps}},\ }\bibfield  {title} {\enquote {\bibinfo
  {title} {{Stimuli-responsive copolymer solution and surface assemblies for
  biomedical applications}},}\ }\href {\doibase 10.1039/c3cs35512h} {\bibfield
  {journal} {\bibinfo  {journal} {Chemical Society Reviews}\ }\textbf {\bibinfo
  {volume} {42}},\ \bibinfo {pages} {7057--7071} (\bibinfo {year}
  {2013})}\BibitemShut {NoStop}%
\bibitem [{\citenamefont {Allen}(2002)}]{Allen2002}%
  \BibitemOpen
  \bibfield  {author} {\bibinfo {author} {\bibfnamefont {Theresa~M.}\
  \bibnamefont {Allen}},\ }\bibfield  {title} {\enquote {\bibinfo {title}
  {{Ligand-targeted therapeutics in anticancer therapy}},}\ }\href {\doibase
  10.1038/nrc903} {\bibfield  {journal} {\bibinfo  {journal} {Nature Reviews
  Cancer 2002 2:10}\ }\textbf {\bibinfo {volume} {2}},\ \bibinfo {pages}
  {750--763} (\bibinfo {year} {2002})}\BibitemShut {NoStop}%
\bibitem [{\citenamefont {Ruiz}\ \emph {et~al.}(2022)\citenamefont {Ruiz},
  \citenamefont {Ramirez},\ and\ \citenamefont {McEnnis}}]{Ruiz2022}%
  \BibitemOpen
  \bibfield  {author} {\bibinfo {author} {\bibfnamefont {Aida~L{\'{o}}pez}\
  \bibnamefont {Ruiz}}, \bibinfo {author} {\bibfnamefont {Ann}\ \bibnamefont
  {Ramirez}}, \ and\ \bibinfo {author} {\bibfnamefont {Kathleen}\ \bibnamefont
  {McEnnis}},\ }\bibfield  {title} {\enquote {\bibinfo {title} {{Single and
  Multiple Stimuli-Responsive Polymer Particles for Controlled Drug
  Delivery}},}\ }\href {\doibase 10.3390/PHARMACEUTICS14020421} {\bibfield
  {journal} {\bibinfo  {journal} {Pharmaceutics 2022, Vol. 14, Page 421}\
  }\textbf {\bibinfo {volume} {14}},\ \bibinfo {pages} {421} (\bibinfo {year}
  {2022})}\BibitemShut {NoStop}%
\bibitem [{\citenamefont {Hadjichristidis}\ \emph {et~al.}(2005)\citenamefont
  {Hadjichristidis}, \citenamefont {Pitsikalis},\ and\ \citenamefont
  {Iatrou}}]{Hadjichristidis2005}%
  \BibitemOpen
  \bibfield  {author} {\bibinfo {author} {\bibfnamefont {Nikos}\ \bibnamefont
  {Hadjichristidis}}, \bibinfo {author} {\bibfnamefont {Marinos}\ \bibnamefont
  {Pitsikalis}}, \ and\ \bibinfo {author} {\bibfnamefont {Hermis}\ \bibnamefont
  {Iatrou}},\ }\bibfield  {title} {\enquote {\bibinfo {title} {{Synthesis of
  Block Copolymers}},}\ }\href {\doibase 10.1007/12_005} {\bibfield  {journal}
  {\bibinfo  {journal} {Advances in Polymer Science}\ }\textbf {\bibinfo
  {volume} {189}},\ \bibinfo {pages} {1--124} (\bibinfo {year}
  {2005})}\BibitemShut {NoStop}%
\bibitem [{\citenamefont {Gitsov}(2008)}]{Gitsov2008}%
  \BibitemOpen
  \bibfield  {author} {\bibinfo {author} {\bibfnamefont {Ivan}\ \bibnamefont
  {Gitsov}},\ }\bibfield  {title} {\enquote {\bibinfo {title} {{Hybrid linear
  dendritic macromolecules: From synthesis to applications}},}\ }\href
  {\doibase 10.1002/POLA.22828} {\bibfield  {journal} {\bibinfo  {journal}
  {Journal of Polymer Science Part A: Polymer Chemistry}\ }\textbf {\bibinfo
  {volume} {46}},\ \bibinfo {pages} {5295--5314} (\bibinfo {year}
  {2008})}\BibitemShut {NoStop}%
\bibitem [{\citenamefont {Wurm}\ and\ \citenamefont {Frey}(2011)}]{Wurm2011}%
  \BibitemOpen
  \bibfield  {author} {\bibinfo {author} {\bibfnamefont {Frederik}\
  \bibnamefont {Wurm}}\ and\ \bibinfo {author} {\bibfnamefont {Holger}\
  \bibnamefont {Frey}},\ }\bibfield  {title} {\enquote {\bibinfo {title}
  {{Linear–dendritic block copolymers: The state of the art and exciting
  perspectives}},}\ }\href {\doibase 10.1016/J.PROGPOLYMSCI.2010.07.009}
  {\bibfield  {journal} {\bibinfo  {journal} {Progress in Polymer Science}\
  }\textbf {\bibinfo {volume} {36}},\ \bibinfo {pages} {1--52} (\bibinfo {year}
  {2011})}\BibitemShut {NoStop}%
\bibitem [{\citenamefont {Fan}\ \emph {et~al.}(2016)\citenamefont {Fan},
  \citenamefont {Zhao}, \citenamefont {Xu},\ and\ \citenamefont
  {Li}}]{Fan2016}%
  \BibitemOpen
  \bibfield  {author} {\bibinfo {author} {\bibfnamefont {Xiaohui}\ \bibnamefont
  {Fan}}, \bibinfo {author} {\bibfnamefont {Yanli}\ \bibnamefont {Zhao}},
  \bibinfo {author} {\bibfnamefont {Wei}\ \bibnamefont {Xu}}, \ and\ \bibinfo
  {author} {\bibfnamefont {Lingbing}\ \bibnamefont {Li}},\ }\bibfield  {title}
  {\enquote {\bibinfo {title} {Linear-dendritic block copolymer for drug and
  gene delivery},}\ }\href {\doibase 10.1016/j.msec.2016.01.044} {\bibfield
  {journal} {\bibinfo  {journal} {Materials Science and Engineering C}\
  }\textbf {\bibinfo {volume} {62}},\ \bibinfo {pages} {943--959} (\bibinfo
  {year} {2016})}\BibitemShut {NoStop}%
\bibitem [{\citenamefont {Whitton}\ and\ \citenamefont
  {Gillies}(2015)}]{Whitton2015}%
  \BibitemOpen
  \bibfield  {author} {\bibinfo {author} {\bibfnamefont {Greg}\ \bibnamefont
  {Whitton}}\ and\ \bibinfo {author} {\bibfnamefont {Elizabeth~R.}\
  \bibnamefont {Gillies}},\ }\bibfield  {title} {\enquote {\bibinfo {title}
  {{Functional aqueous assemblies of linear-dendron hybrids}},}\ }\href
  {\doibase 10.1002/POLA.27316} {\bibfield  {journal} {\bibinfo  {journal}
  {Journal of Polymer Science, Part A: Polymer Chemistry}\ }\textbf {\bibinfo
  {volume} {53}},\ \bibinfo {pages} {148--172} (\bibinfo {year}
  {2015})}\BibitemShut {NoStop}%
\bibitem [{\citenamefont {Liu}\ and\ \citenamefont {Gitsov}(2019)}]{Liu2019}%
  \BibitemOpen
  \bibfield  {author} {\bibinfo {author} {\bibfnamefont {Xin}\ \bibnamefont
  {Liu}}\ and\ \bibinfo {author} {\bibfnamefont {Ivan}\ \bibnamefont
  {Gitsov}},\ }\bibfield  {title} {\enquote {\bibinfo {title} {{Nonionic
  Amphiphilic Linear Dendritic Block Copolymers. Solvent-Induced Self-Assembly
  and Morphology Tuning}},}\ }\href {\doibase
  10.1021/ACS.MACROMOL.9B01023/ASSET/IMAGES/LARGE/MA-2019-01023K_0004.JPEG}
  {\bibfield  {journal} {\bibinfo  {journal} {Macromolecules}\ }\textbf
  {\bibinfo {volume} {52}},\ \bibinfo {pages} {5563--5573} (\bibinfo {year}
  {2019})}\BibitemShut {NoStop}%
\bibitem [{\citenamefont {Lebedeva}\ \emph {et~al.}(2018)\citenamefont
  {Lebedeva}, \citenamefont {Zhulina},\ and\ \citenamefont
  {Borisov}}]{Lebedeva2018}%
  \BibitemOpen
  \bibfield  {author} {\bibinfo {author} {\bibfnamefont {Inna~O.}\ \bibnamefont
  {Lebedeva}}, \bibinfo {author} {\bibfnamefont {Ekaterina~B.}\ \bibnamefont
  {Zhulina}}, \ and\ \bibinfo {author} {\bibfnamefont {Oleg~V.}\ \bibnamefont
  {Borisov}},\ }\bibfield  {title} {\enquote {\bibinfo {title} {{Theory of
  Linear-Dendritic Block Copolymer Micelles}},}\ }\href {\doibase
  10.1021/ACSMACROLETT.7B00784/ASSET/IMAGES/LARGE/MZ-2017-007843_0003.JPEG}
  {\bibfield  {journal} {\bibinfo  {journal} {ACS Macro Letters}\ }\textbf
  {\bibinfo {volume} {7}},\ \bibinfo {pages} {42--46} (\bibinfo {year}
  {2018})}\BibitemShut {NoStop}%
\bibitem [{\citenamefont {Tomalia}\ \emph {et~al.}(1985)\citenamefont
  {Tomalia}, \citenamefont {Baker}, \citenamefont {Dewald}, \citenamefont
  {Hall}, \citenamefont {Kallos}, \citenamefont {Martin}, \citenamefont
  {Roeck}, \citenamefont {Ryder},\ and\ \citenamefont {Smith}}]{Tomalia1985}%
  \BibitemOpen
  \bibfield  {author} {\bibinfo {author} {\bibfnamefont {DA}~\bibnamefont
  {Tomalia}}, \bibinfo {author} {\bibfnamefont {H}~\bibnamefont {Baker}},
  \bibinfo {author} {\bibfnamefont {J}~\bibnamefont {Dewald}}, \bibinfo
  {author} {\bibfnamefont {M}~\bibnamefont {Hall}}, \bibinfo {author}
  {\bibfnamefont {G}~\bibnamefont {Kallos}}, \bibinfo {author} {\bibfnamefont
  {S}~\bibnamefont {Martin}}, \bibinfo {author} {\bibfnamefont {J}~\bibnamefont
  {Roeck}}, \bibinfo {author} {\bibfnamefont {J}~\bibnamefont {Ryder}}, \ and\
  \bibinfo {author} {\bibfnamefont {P}~\bibnamefont {Smith}},\ }\bibfield
  {title} {\enquote {\bibinfo {title} {A new class of polymers:
  Starburst-dendritic macromolecules},}\ }\href {\doibase
  10.1295/polymj.17.117} {\bibfield  {journal} {\bibinfo  {journal} {Polymer
  Journal}\ }\textbf {\bibinfo {volume} {17}},\ \bibinfo {pages} {117--132}
  (\bibinfo {year} {1985})}\BibitemShut {NoStop}%
\bibitem [{\citenamefont {Grayson}\ and\ \citenamefont
  {Fr{\'{e}}chet}(2001)}]{Grayson2001}%
  \BibitemOpen
  \bibfield  {author} {\bibinfo {author} {\bibfnamefont {Scott~M.}\
  \bibnamefont {Grayson}}\ and\ \bibinfo {author} {\bibfnamefont {Jean~M.J.}\
  \bibnamefont {Fr{\'{e}}chet}},\ }\bibfield  {title} {\enquote {\bibinfo
  {title} {{Divergent synthesis of dendronized poly(p-hydroxystyrene) [6]}},}\
  }\href {\doibase 10.1021/MA010698G/SUPPL_FILE/MA010698G_S.PDF} {\bibfield
  {journal} {\bibinfo  {journal} {Macromolecules}\ }\textbf {\bibinfo {volume}
  {34}},\ \bibinfo {pages} {6542--6544} (\bibinfo {year} {2001})}\BibitemShut
  {NoStop}%
\bibitem [{\citenamefont {Nuhn}\ \emph {et~al.}(2013)\citenamefont {Nuhn},
  \citenamefont {Sch{\"{u}}ll}, \citenamefont {Frey},\ and\ \citenamefont
  {Zentel}}]{Nuhn2013}%
  \BibitemOpen
  \bibfield  {author} {\bibinfo {author} {\bibfnamefont {Lutz}\ \bibnamefont
  {Nuhn}}, \bibinfo {author} {\bibfnamefont {Christoph}\ \bibnamefont
  {Sch{\"{u}}ll}}, \bibinfo {author} {\bibfnamefont {Holger}\ \bibnamefont
  {Frey}}, \ and\ \bibinfo {author} {\bibfnamefont {Rudolf}\ \bibnamefont
  {Zentel}},\ }\bibfield  {title} {\enquote {\bibinfo {title} {{Combining
  ring-opening multibranching and RAFT polymerization: Multifunctional
  linear-hyperbranched block copolymers via hyperbranched macro-chain-transfer
  agents}},}\ }\href {\doibase
  10.1021/MA4002897/SUPPL_FILE/MA4002897_SI_001.PDF} {\bibfield  {journal}
  {\bibinfo  {journal} {Macromolecules}\ }\textbf {\bibinfo {volume} {46}},\
  \bibinfo {pages} {2892--2904} (\bibinfo {year} {2013})}\BibitemShut {NoStop}%
\bibitem [{\citenamefont {Oikawa}\ \emph {et~al.}(2013)\citenamefont {Oikawa},
  \citenamefont {Lee}, \citenamefont {Kim}, \citenamefont {Kang}, \citenamefont
  {Kim}, \citenamefont {Saito}, \citenamefont {Sasaki}, \citenamefont {Oishi},\
  and\ \citenamefont {Shibasaki}}]{oikawa2013one}%
  \BibitemOpen
  \bibfield  {author} {\bibinfo {author} {\bibfnamefont {Yurie}\ \bibnamefont
  {Oikawa}}, \bibinfo {author} {\bibfnamefont {Sueun}\ \bibnamefont {Lee}},
  \bibinfo {author} {\bibfnamefont {Do~Hyung}\ \bibnamefont {Kim}}, \bibinfo
  {author} {\bibfnamefont {Dae~Hwan}\ \bibnamefont {Kang}}, \bibinfo {author}
  {\bibfnamefont {Byeong-Su}\ \bibnamefont {Kim}}, \bibinfo {author}
  {\bibfnamefont {Kyohei}\ \bibnamefont {Saito}}, \bibinfo {author}
  {\bibfnamefont {Shigeko}\ \bibnamefont {Sasaki}}, \bibinfo {author}
  {\bibfnamefont {Yoshiyuki}\ \bibnamefont {Oishi}}, \ and\ \bibinfo {author}
  {\bibfnamefont {Yuji}\ \bibnamefont {Shibasaki}},\ }\bibfield  {title}
  {\enquote {\bibinfo {title} {One-pot synthesis of linear-hyperbranched
  amphiphilic block copolymers based on polyglycerol derivatives and their
  micelles},}\ }\href {\doibase 10.1021/bm400275w} {\bibfield  {journal}
  {\bibinfo  {journal} {Biomacromolecules}\ }\textbf {\bibinfo {volume} {14}},\
  \bibinfo {pages} {2171--2178} (\bibinfo {year} {2013})}\BibitemShut {NoStop}%
\bibitem [{\citenamefont {Cuneo}\ and\ \citenamefont {Gao}(2020)}]{Cuneo2020}%
  \BibitemOpen
  \bibfield  {author} {\bibinfo {author} {\bibfnamefont {Timothy}\ \bibnamefont
  {Cuneo}}\ and\ \bibinfo {author} {\bibfnamefont {Haifeng}\ \bibnamefont
  {Gao}},\ }\bibfield  {title} {\enquote {\bibinfo {title} {Recent advances on
  synthesis and biomaterials applications of hyperbranched polymers},}\ }\href
  {https://doi.org/10.1002/wnan.1640} {\bibfield  {journal} {\bibinfo
  {journal} {WRES - Nanomedicine and Nanobiotechnology}\ }\textbf {\bibinfo
  {volume} {12}} (\bibinfo {year} {2020})}\BibitemShut {NoStop}%
\bibitem [{\citenamefont {Uhrich}\ \emph {et~al.}(1992)\citenamefont {Uhrich},
  \citenamefont {Hawker}, \citenamefont {Fr{\'{e}}chet},\ and\ \citenamefont
  {Turner}}]{Uhrich1992}%
  \BibitemOpen
  \bibfield  {author} {\bibinfo {author} {\bibfnamefont {K~E}\ \bibnamefont
  {Uhrich}}, \bibinfo {author} {\bibfnamefont {C~J}\ \bibnamefont {Hawker}},
  \bibinfo {author} {\bibfnamefont {J~M~J}\ \bibnamefont {Fr{\'{e}}chet}}, \
  and\ \bibinfo {author} {\bibfnamefont {S~R}\ \bibnamefont {Turner}},\
  }\bibfield  {title} {\enquote {\bibinfo {title} {{One-Pot Synthesis of
  Hyperbranched Polyethers}},}\ }\href {https://pubs.acs.org/sharingguidelines}
  {\bibfield  {journal} {\bibinfo  {journal} {Macromolecules}\ }\textbf
  {\bibinfo {volume} {25}},\ \bibinfo {pages} {4583--4587} (\bibinfo {year}
  {1992})}\BibitemShut {NoStop}%
\bibitem [{\citenamefont {Kim}\ and\ \citenamefont {Webster}(1990)}]{Kim1990}%
  \BibitemOpen
  \bibfield  {author} {\bibinfo {author} {\bibfnamefont {Young~H}\ \bibnamefont
  {Kim}}\ and\ \bibinfo {author} {\bibfnamefont {Owen~W}\ \bibnamefont
  {Webster}},\ }\bibfield  {title} {\enquote {\bibinfo {title} {Water soluble
  hyperbranched polyphenylene:" a unimolecular micelle?"},}\ }\href {\doibase
  10.1021/ja00167a094} {\bibfield  {journal} {\bibinfo  {journal} {Journal of
  the American Chemical Society}\ }\textbf {\bibinfo {volume} {112}},\ \bibinfo
  {pages} {4592--4593} (\bibinfo {year} {1990})}\BibitemShut {NoStop}%
\bibitem [{\citenamefont {Andresen}\ and\ \citenamefont
  {Larsen}(2020)}]{Andresen2020}%
  \BibitemOpen
  \bibfield  {author} {\bibinfo {author} {\bibfnamefont {Thomas~Lars}\
  \bibnamefont {Andresen}}\ and\ \bibinfo {author} {\bibfnamefont
  {Jannik~Bruun}\ \bibnamefont {Larsen}},\ }\bibfield  {title} {\enquote
  {\bibinfo {title} {{Compositional inhomogeneity of drug delivery liposomes
  quantified at the single liposome level}},}\ }\href {\doibase
  10.1016/J.ACTBIO.2020.10.003} {\bibfield  {journal} {\bibinfo  {journal}
  {Acta Biomaterialia}\ }\textbf {\bibinfo {volume} {118}},\ \bibinfo {pages}
  {207--214} (\bibinfo {year} {2020})}\BibitemShut {NoStop}%
\bibitem [{\citenamefont {Schmitt}\ \emph {et~al.}(2012)\citenamefont
  {Schmitt}, \citenamefont {Repollet-Pedrosa},\ and\ \citenamefont
  {Mahanthappa}}]{Schmitt2012}%
  \BibitemOpen
  \bibfield  {author} {\bibinfo {author} {\bibfnamefont {Andrew~L.}\
  \bibnamefont {Schmitt}}, \bibinfo {author} {\bibfnamefont {Milton~H.}\
  \bibnamefont {Repollet-Pedrosa}}, \ and\ \bibinfo {author} {\bibfnamefont
  {Mahesh~K.}\ \bibnamefont {Mahanthappa}},\ }\bibfield  {title} {\enquote
  {\bibinfo {title} {{Polydispersity-driven block copolymer amphiphile
  self-assembly into prolate-spheroid micelles}},}\ }\href {\doibase
  10.1021/MZ200156S/SUPPL_FILE/MZ200156S_SI_001.PDF} {\bibfield  {journal}
  {\bibinfo  {journal} {ACS Macro Letters}\ }\textbf {\bibinfo {volume} {1}},\
  \bibinfo {pages} {300--304} (\bibinfo {year} {2012})}\BibitemShut {NoStop}%
\bibitem [{\citenamefont {Barriau}\ \emph {et~al.}(2005)\citenamefont
  {Barriau}, \citenamefont {{Garc{\'{i}}a Marcos}}, \citenamefont {Kautz},\
  and\ \citenamefont {Frey}}]{Barriau2005}%
  \BibitemOpen
  \bibfield  {author} {\bibinfo {author} {\bibfnamefont {Emilie}\ \bibnamefont
  {Barriau}}, \bibinfo {author} {\bibfnamefont {Alejandra}\ \bibnamefont
  {{Garc{\'{i}}a Marcos}}}, \bibinfo {author} {\bibfnamefont {Holger}\
  \bibnamefont {Kautz}}, \ and\ \bibinfo {author} {\bibfnamefont {Holger}\
  \bibnamefont {Frey}},\ }\bibfield  {title} {\enquote {\bibinfo {title}
  {{Linear-Hyperbranched Amphiphilic AB Diblock Copolymers Based on Polystyrene
  and Hyperbranched Polyglycerol}},}\ }\href {\doibase 10.1002/MARC.200500184}
  {\bibfield  {journal} {\bibinfo  {journal} {Macromolecular Rapid
  Communications}\ }\textbf {\bibinfo {volume} {26}},\ \bibinfo {pages}
  {862--867} (\bibinfo {year} {2005})}\BibitemShut {NoStop}%
\bibitem [{\citenamefont {Mantha}\ \emph {et~al.}(2019)\citenamefont {Mantha},
  \citenamefont {Qi}, \citenamefont {Barz},\ and\ \citenamefont
  {Schmid}}]{Mantha2019}%
  \BibitemOpen
  \bibfield  {author} {\bibinfo {author} {\bibfnamefont {Sriteja}\ \bibnamefont
  {Mantha}}, \bibinfo {author} {\bibfnamefont {Shuanhu}\ \bibnamefont {Qi}},
  \bibinfo {author} {\bibfnamefont {Matthias}\ \bibnamefont {Barz}}, \ and\
  \bibinfo {author} {\bibfnamefont {Friederike}\ \bibnamefont {Schmid}},\
  }\bibfield  {title} {\enquote {\bibinfo {title} {How ill-defined constituents
  produce well-defined nanoparticles: Effect of polymer dispersity on the
  uniformity of copolymeric micelles},}\ }\href {\doibase
  10.1103/PhysRevMaterials.3.026002} {\bibfield  {journal} {\bibinfo  {journal}
  {Physical Review Materials}\ }\textbf {\bibinfo {volume} {3}},\ \bibinfo
  {pages} {026002} (\bibinfo {year} {2019})}\BibitemShut {NoStop}%
\bibitem [{\citenamefont {{DE GENNES}}(1978)}]{DEGENNES1978}%
  \BibitemOpen
  \bibfield  {author} {\bibinfo {author} {\bibfnamefont {P.G.}\ \bibnamefont
  {{DE GENNES}}},\ }\bibfield  {title} {\enquote {\bibinfo {title}
  {{Macromolecules and Liquid Crystals: Reflections on Certain Lines of
  Research}},}\ }\href {\doibase 10.1016/B978-0-12-607774-2.50006-9} {\bibfield
   {journal} {\bibinfo  {journal} {Liquid Crystals}\ ,\ \bibinfo {pages}
  {1--18}} (\bibinfo {year} {1978})}\BibitemShut {NoStop}%
\bibitem [{\citenamefont {Noolandi}\ and\ \citenamefont
  {Hong}(1983)}]{Noolandi1983}%
  \BibitemOpen
  \bibfield  {author} {\bibinfo {author} {\bibfnamefont {Jaan}\ \bibnamefont
  {Noolandi}}\ and\ \bibinfo {author} {\bibfnamefont {Kin~Ming}\ \bibnamefont
  {Hong}},\ }\bibfield  {title} {\enquote {\bibinfo {title} {{Theory of Block
  Copolymer Micelles in Solution}},}\ }\href {\doibase 10.1021/ma00243a007}
  {\bibfield  {journal} {\bibinfo  {journal} {Macromolecules}\ }\textbf
  {\bibinfo {volume} {16}},\ \bibinfo {pages} {1443--1448} (\bibinfo {year}
  {1983})}\BibitemShut {NoStop}%
\bibitem [{\citenamefont {Leibler}\ \emph {et~al.}(1983)\citenamefont
  {Leibler}, \citenamefont {Orland},\ and\ \citenamefont
  {Wheeler}}]{Leibler1983}%
  \BibitemOpen
  \bibfield  {author} {\bibinfo {author} {\bibfnamefont {Ludwik}\ \bibnamefont
  {Leibler}}, \bibinfo {author} {\bibfnamefont {Henri}\ \bibnamefont {Orland}},
  \ and\ \bibinfo {author} {\bibfnamefont {John~C.}\ \bibnamefont {Wheeler}},\
  }\bibfield  {title} {\enquote {\bibinfo {title} {{Theory of critical micelle
  concentration for solutions of block copolymers}},}\ }\href {\doibase
  10.1063/1.446209} {\bibfield  {journal} {\bibinfo  {journal} {The Journal of
  Chemical Physics}\ }\textbf {\bibinfo {volume} {79}},\ \bibinfo {pages}
  {3550--3557} (\bibinfo {year} {1983})}\BibitemShut {NoStop}%
\bibitem [{\citenamefont {Leermakers}\ \emph {et~al.}(1995)\citenamefont
  {Leermakers}, \citenamefont {Wijmans},\ and\ \citenamefont
  {Fleer}}]{Leermakers1995}%
  \BibitemOpen
  \bibfield  {author} {\bibinfo {author} {\bibfnamefont {F.~A.~M.}\
  \bibnamefont {Leermakers}}, \bibinfo {author} {\bibfnamefont {C.~M.}\
  \bibnamefont {Wijmans}}, \ and\ \bibinfo {author} {\bibfnamefont {G.~J.}\
  \bibnamefont {Fleer}},\ }\bibfield  {title} {\enquote {\bibinfo {title} {On
  the structure of polymeric micelles: Self-consistent-field theory and
  universal properties for volume fraction profiles},}\ }\href {\doibase
  10.1021/ma00113a050} {\bibfield  {journal} {\bibinfo  {journal}
  {Macromolecules}\ }\textbf {\bibinfo {volume} {28}},\ \bibinfo {pages}
  {3434--3443} (\bibinfo {year} {1995})}\BibitemShut {NoStop}%
\bibitem [{\citenamefont {Nelson}\ \emph {et~al.}(1997)\citenamefont {Nelson},
  \citenamefont {Rutledge},\ and\ \citenamefont {Hatton}}]{Nelson1997}%
  \BibitemOpen
  \bibfield  {author} {\bibinfo {author} {\bibfnamefont {P.~H.}\ \bibnamefont
  {Nelson}}, \bibinfo {author} {\bibfnamefont {G.~C.}\ \bibnamefont
  {Rutledge}}, \ and\ \bibinfo {author} {\bibfnamefont {T.~A.}\ \bibnamefont
  {Hatton}},\ }\bibfield  {title} {\enquote {\bibinfo {title} {On the size and
  shape of self-assembled micelles},}\ }\href {\doibase 10.1063/1.474193}
  {\bibfield  {journal} {\bibinfo  {journal} {J. Chem. Phys.}\ }\textbf
  {\bibinfo {volume} {107}},\ \bibinfo {pages} {10777--10781} (\bibinfo {year}
  {1997})}\BibitemShut {NoStop}%
\bibitem [{\citenamefont {Zhulina}\ and\ \citenamefont
  {Borisov}(2012)}]{zhulina_theory_2012}%
  \BibitemOpen
  \bibfield  {author} {\bibinfo {author} {\bibfnamefont {E.~B.}\ \bibnamefont
  {Zhulina}}\ and\ \bibinfo {author} {\bibfnamefont {O.~V.}\ \bibnamefont
  {Borisov}},\ }\bibfield  {title} {\enquote {\bibinfo {title} {Theory of
  {Block} {Polymer} {Micelles}: {Recent} {Advances} and {Current}
  {Challenges}},}\ }\href {\doibase 10.1021/ma300195n} {\bibfield  {journal}
  {\bibinfo  {journal} {Macromolecules}\ }\textbf {\bibinfo {volume} {45}},\
  \bibinfo {pages} {4429--4440} (\bibinfo {year} {2012})}\BibitemShut {NoStop}%
\bibitem [{\citenamefont {Wang}\ \emph {et~al.}(2013)\citenamefont {Wang},
  \citenamefont {Li}, \citenamefont {Zhou}, \citenamefont {Lu},\ and\
  \citenamefont {Yan}}]{Wang2013}%
  \BibitemOpen
  \bibfield  {author} {\bibinfo {author} {\bibfnamefont {Yuling}\ \bibnamefont
  {Wang}}, \bibinfo {author} {\bibfnamefont {Bin}\ \bibnamefont {Li}}, \bibinfo
  {author} {\bibfnamefont {Yongfeng}\ \bibnamefont {Zhou}}, \bibinfo {author}
  {\bibfnamefont {Zhongyuan}\ \bibnamefont {Lu}}, \ and\ \bibinfo {author}
  {\bibfnamefont {Deyue}\ \bibnamefont {Yan}},\ }\bibfield  {title} {\enquote
  {\bibinfo {title} {Dissipative particle dynamics simulation study on the
  mechanisms of self-assembly of large multimolecular micelles from amphiphilic
  dendritic multiarm copolymers},}\ }\href {\doibase 10.1039/c3sm27396b}
  {\bibfield  {journal} {\bibinfo  {journal} {Soft Matter}\ }\textbf {\bibinfo
  {volume} {9}},\ \bibinfo {pages} {3293--3304} (\bibinfo {year}
  {2013})}\BibitemShut {NoStop}%
\bibitem [{\citenamefont {Lebedeva}\ \emph {et~al.}(2019)\citenamefont
  {Lebedeva}, \citenamefont {Zhulina},\ and\ \citenamefont
  {Borisov}}]{lebedeva_self-assembly_2019}%
  \BibitemOpen
  \bibfield  {author} {\bibinfo {author} {\bibfnamefont {Inna~O.}\ \bibnamefont
  {Lebedeva}}, \bibinfo {author} {\bibfnamefont {Ekaterina~B.}\ \bibnamefont
  {Zhulina}}, \ and\ \bibinfo {author} {\bibfnamefont {Oleg~V.}\ \bibnamefont
  {Borisov}},\ }\bibfield  {title} {\enquote {\bibinfo {title} {Self-{Assembly}
  of {Linear}-{Dendritic} and {Double} {Dendritic} {Block} {Copolymers}: {From}
  {Dendromicelles} to {Dendrimersomes}},}\ }\href {\doibase
  10.1021/acs.macromol.9b00140} {\bibfield  {journal} {\bibinfo  {journal}
  {Macromolecules}\ }\textbf {\bibinfo {volume} {52}},\ \bibinfo {pages}
  {3655--3667} (\bibinfo {year} {2019})}\BibitemShut {NoStop}%
\bibitem [{\citenamefont {Brito}\ \emph {et~al.}(2023)\citenamefont {Brito},
  \citenamefont {Mikhtaniuk}, \citenamefont {Neelov}, \citenamefont {Borisov},\
  and\ \citenamefont {Holm}}]{Brito2023}%
  \BibitemOpen
  \bibfield  {author} {\bibinfo {author} {\bibfnamefont {Mariano~E.}\
  \bibnamefont {Brito}}, \bibinfo {author} {\bibfnamefont {Sofia~E.}\
  \bibnamefont {Mikhtaniuk}}, \bibinfo {author} {\bibfnamefont {Igor~M.}\
  \bibnamefont {Neelov}}, \bibinfo {author} {\bibfnamefont {Oleg~V.}\
  \bibnamefont {Borisov}}, \ and\ \bibinfo {author} {\bibfnamefont {Christian}\
  \bibnamefont {Holm}},\ }\bibfield  {title} {\enquote {\bibinfo {title}
  {{Implicit-Solvent Coarse-Grained Simulations of Linear–Dendritic Block
  Copolymer Micelles}},}\ }\href {\doibase 10.3390/IJMS24032763} {\bibfield
  {journal} {\bibinfo  {journal} {International Journal of Molecular Sciences}\
  }\textbf {\bibinfo {volume} {24}},\ \bibinfo {pages} {2763} (\bibinfo {year}
  {2023})}\BibitemShut {NoStop}%
\bibitem [{\citenamefont {Tan}\ \emph {et~al.}(2015)\citenamefont {Tan},
  \citenamefont {Wang}, \citenamefont {Yu}, \citenamefont {Zhou}, \citenamefont
  {Lu},\ and\ \citenamefont {Yan}}]{Tan2015}%
  \BibitemOpen
  \bibfield  {author} {\bibinfo {author} {\bibfnamefont {Haina}\ \bibnamefont
  {Tan}}, \bibinfo {author} {\bibfnamefont {Wei}\ \bibnamefont {Wang}},
  \bibinfo {author} {\bibfnamefont {Chunyang}\ \bibnamefont {Yu}}, \bibinfo
  {author} {\bibfnamefont {Yongfeng}\ \bibnamefont {Zhou}}, \bibinfo {author}
  {\bibfnamefont {Zhongyuan}\ \bibnamefont {Lu}}, \ and\ \bibinfo {author}
  {\bibfnamefont {Deyue}\ \bibnamefont {Yan}},\ }\bibfield  {title} {\enquote
  {\bibinfo {title} {Dissipative particle dynamics simulation study on
  self-assembly of amphiphilic hyperbranched multiarm copolymers with different
  degrees of branching},}\ }\href {\doibase https://doi.org/10.1039/C5SM01495F}
  {\bibfield  {journal} {\bibinfo  {journal} {Soft Matter}\ }\textbf {\bibinfo
  {volume} {11}},\ \bibinfo {pages} {8460--8470} (\bibinfo {year}
  {2015})}\BibitemShut {NoStop}%
\bibitem [{\citenamefont {Tan}\ \emph {et~al.}(2017)\citenamefont {Tan},
  \citenamefont {Yu}, \citenamefont {Lu}, \citenamefont {Zhou},\ and\
  \citenamefont {Yan}}]{Tan2017}%
  \BibitemOpen
  \bibfield  {author} {\bibinfo {author} {\bibfnamefont {Haina}\ \bibnamefont
  {Tan}}, \bibinfo {author} {\bibfnamefont {Chunyang}\ \bibnamefont {Yu}},
  \bibinfo {author} {\bibfnamefont {Zhongyuan}\ \bibnamefont {Lu}}, \bibinfo
  {author} {\bibfnamefont {Yongfeng}\ \bibnamefont {Zhou}}, \ and\ \bibinfo
  {author} {\bibfnamefont {Deyue}\ \bibnamefont {Yan}},\ }\bibfield  {title}
  {\enquote {\bibinfo {title} {A dissipative particle dynamics simulation study
  on phase diagrams for the self-assembly of amphiphilic hyperbranched multiarm
  copolymers in various solvents},}\ }\href {\doibase 10.1039/c7sm01170a}
  {\bibfield  {journal} {\bibinfo  {journal} {Soft Matter}\ }\textbf {\bibinfo
  {volume} {13}},\ \bibinfo {pages} {6178--6188} (\bibinfo {year}
  {2017})}\BibitemShut {NoStop}%
\bibitem [{\citenamefont {Tan}\ \emph {et~al.}(2019)\citenamefont {Tan},
  \citenamefont {Li}, \citenamefont {Li}, \citenamefont {Yu}, \citenamefont
  {Lu},\ and\ \citenamefont {Zhou}}]{Tan2019}%
  \BibitemOpen
  \bibfield  {author} {\bibinfo {author} {\bibfnamefont {Haina}\ \bibnamefont
  {Tan}}, \bibinfo {author} {\bibfnamefont {Shanlong}\ \bibnamefont {Li}},
  \bibinfo {author} {\bibfnamefont {Ke}~\bibnamefont {Li}}, \bibinfo {author}
  {\bibfnamefont {Chunyang}\ \bibnamefont {Yu}}, \bibinfo {author}
  {\bibfnamefont {Zhongyuan}\ \bibnamefont {Lu}}, \ and\ \bibinfo {author}
  {\bibfnamefont {Yongfeng}\ \bibnamefont {Zhou}},\ }\bibfield  {title}
  {\enquote {\bibinfo {title} {Shape transformations of vesicles self-assembled
  from amphiphilic hyperbranched multiarm copolymers via simulation},}\ }\href
  {\doibase 10.1021/acs.langmuir.8b02206} {\bibfield  {journal} {\bibinfo
  {journal} {Langmuir}\ }\textbf {\bibinfo {volume} {35}},\ \bibinfo {pages}
  {6929--6938} (\bibinfo {year} {2019})}\BibitemShut {NoStop}%
\bibitem [{\citenamefont {Hao}\ \emph {et~al.}(2020)\citenamefont {Hao},
  \citenamefont {Tan}, \citenamefont {Li}, \citenamefont {Wang}, \citenamefont
  {Zhou}, \citenamefont {Yu}, \citenamefont {Zhou},\ and\ \citenamefont
  {Yan}}]{Hao2020}%
  \BibitemOpen
  \bibfield  {author} {\bibinfo {author} {\bibfnamefont {Tongfan}\ \bibnamefont
  {Hao}}, \bibinfo {author} {\bibfnamefont {Haina}\ \bibnamefont {Tan}},
  \bibinfo {author} {\bibfnamefont {Shanlong}\ \bibnamefont {Li}}, \bibinfo
  {author} {\bibfnamefont {Yuling}\ \bibnamefont {Wang}}, \bibinfo {author}
  {\bibfnamefont {Zhiping}\ \bibnamefont {Zhou}}, \bibinfo {author}
  {\bibfnamefont {Chunyang}\ \bibnamefont {Yu}}, \bibinfo {author}
  {\bibfnamefont {Yongfeng}\ \bibnamefont {Zhou}}, \ and\ \bibinfo {author}
  {\bibfnamefont {Deyue}\ \bibnamefont {Yan}},\ }\bibfield  {title} {\enquote
  {\bibinfo {title} {Multilayer onion-like vesicles self-assembled from
  amphiphilic hyperbranched multiarm copolymers via simulation},}\ }\href
  {\doibase 10.1002/pol.20190163} {\bibfield  {journal} {\bibinfo  {journal}
  {J. Pol. Science}\ }\textbf {\bibinfo {volume} {58}},\ \bibinfo {pages}
  {704--715} (\bibinfo {year} {2020})}\BibitemShut {NoStop}%
\bibitem [{\citenamefont {Gao}\ and\ \citenamefont
  {Eisenberg}(1993)}]{Gao1993}%
  \BibitemOpen
  \bibfield  {author} {\bibinfo {author} {\bibfnamefont {Zhisheng}\
  \bibnamefont {Gao}}\ and\ \bibinfo {author} {\bibfnamefont {Adi}\
  \bibnamefont {Eisenberg}},\ }\bibfield  {title} {\enquote {\bibinfo {title}
  {A model of micellization for block copolymers in solutions},}\ }\href
  {\doibase 10.1021/ma00078a035} {\bibfield  {journal} {\bibinfo  {journal}
  {Macromolecules}\ }\textbf {\bibinfo {volume} {26}},\ \bibinfo {pages}
  {7353--7360} (\bibinfo {year} {1993})}\BibitemShut {NoStop}%
\bibitem [{\citenamefont {Linse}(1994)}]{Linse1994}%
  \BibitemOpen
  \bibfield  {author} {\bibinfo {author} {\bibfnamefont {Per}\ \bibnamefont
  {Linse}},\ }\bibfield  {title} {\enquote {\bibinfo {title} {Micellization of
  poly(ethylene oxide)--poly(propylene oxide) block copolymers in aqueous
  solution: Effect of polymer polydispersity},}\ }\href {\doibase
  10.1021/ma00069a007} {\bibfield  {journal} {\bibinfo  {journal}
  {Macromolecules}\ }\textbf {\bibinfo {volume} {27}},\ \bibinfo {pages}
  {6404--6417} (\bibinfo {year} {1994})}\BibitemShut {NoStop}%
\bibitem [{\citenamefont {Lynd}\ \emph {et~al.}(2008)\citenamefont {Lynd},
  \citenamefont {Meuler},\ and\ \citenamefont {Hillmyer}}]{Lynd2008}%
  \BibitemOpen
  \bibfield  {author} {\bibinfo {author} {\bibfnamefont {Nathaniel~A.}\
  \bibnamefont {Lynd}}, \bibinfo {author} {\bibfnamefont {Adam~J.}\
  \bibnamefont {Meuler}}, \ and\ \bibinfo {author} {\bibfnamefont {Marc~A.}\
  \bibnamefont {Hillmyer}},\ }\bibfield  {title} {\enquote {\bibinfo {title}
  {Polydispersity and block copolymer self-assembly},}\ }\href {\doibase
  https://doi.org/10.1016/j.progpolymsci.2008.07.003} {\bibfield  {journal}
  {\bibinfo  {journal} {Prog. Polym. Sci.}\ }\textbf {\bibinfo {volume} {33}},\
  \bibinfo {pages} {875 -- 893} (\bibinfo {year} {2008})}\BibitemShut {NoStop}%
\bibitem [{\citenamefont {Doncom}\ \emph {et~al.}(2017)\citenamefont {Doncom},
  \citenamefont {Blackman}, \citenamefont {Wright}, \citenamefont {Gibson},\
  and\ \citenamefont {O{'}Reilly}}]{Doncom2017}%
  \BibitemOpen
  \bibfield  {author} {\bibinfo {author} {\bibfnamefont {Kay E.~B.}\
  \bibnamefont {Doncom}}, \bibinfo {author} {\bibfnamefont {Lewis~D.}\
  \bibnamefont {Blackman}}, \bibinfo {author} {\bibfnamefont {Daniel~B.}\
  \bibnamefont {Wright}}, \bibinfo {author} {\bibfnamefont {Matthew~I.}\
  \bibnamefont {Gibson}}, \ and\ \bibinfo {author} {\bibfnamefont {Rachel~K.}\
  \bibnamefont {O{'}Reilly}},\ }\bibfield  {title} {\enquote {\bibinfo {title}
  {Dispersity effects in polymer self-assemblies: a matter of hierarchical
  control},}\ }\href {\doibase 10.1039/C6CS00818F} {\bibfield  {journal}
  {\bibinfo  {journal} {Chem. Soc. Rev.}\ }\textbf {\bibinfo {volume} {46}},\
  \bibinfo {pages} {4119--4134} (\bibinfo {year} {2017})}\BibitemShut {NoStop}%
\bibitem [{\citenamefont {Giannakou}\ \emph {et~al.}(2024)\citenamefont
  {Giannakou}, \citenamefont {Borisov},\ and\ \citenamefont
  {Schmid}}]{Giannakou2024}%
  \BibitemOpen
  \bibfield  {author} {\bibinfo {author} {\bibfnamefont {Marios}\ \bibnamefont
  {Giannakou}}, \bibinfo {author} {\bibfnamefont {Oleg~V}\ \bibnamefont
  {Borisov}}, \ and\ \bibinfo {author} {\bibfnamefont {Friederike}\
  \bibnamefont {Schmid}},\ }\bibfield  {title} {\enquote {\bibinfo {title}
  {Strong stretching theory of polydisperse curved polymer brushes},}\ }\href
  {\doibase 10.1063/5.0213524} {\bibfield  {journal} {\bibinfo  {journal} {The
  Journal of Chemical Physics}\ }\textbf {\bibinfo {volume} {161}} (\bibinfo
  {year} {2024}),\ 10.1063/5.0213524}\BibitemShut {NoStop}%
\bibitem [{\citenamefont {Sunder}\ \emph {et~al.}(1999)\citenamefont {Sunder},
  \citenamefont {Hanselmann}, \citenamefont {Frey},\ and\ \citenamefont
  {Mülhaupt}}]{sunder1999controlled}%
  \BibitemOpen
  \bibfield  {author} {\bibinfo {author} {\bibfnamefont {A}~\bibnamefont
  {Sunder}}, \bibinfo {author} {\bibfnamefont {R}~\bibnamefont {Hanselmann}},
  \bibinfo {author} {\bibfnamefont {H}~\bibnamefont {Frey}}, \ and\ \bibinfo
  {author} {\bibfnamefont {R}~\bibnamefont {Mülhaupt}},\ }\bibfield  {title}
  {\enquote {\bibinfo {title} {Controlled synthesis of hyperbranched
  polyglycerols by ring-opening multibranching polymerization},}\ }\href
  {\doibase 10.1021/ma990090w} {\bibfield  {journal} {\bibinfo  {journal}
  {Macromolecules}\ }\textbf {\bibinfo {volume} {32}},\ \bibinfo {pages}
  {4240--4246} (\bibinfo {year} {1999})}\BibitemShut {NoStop}%
\bibitem [{\citenamefont {Schuell}\ \emph {et~al.}(2013)\citenamefont
  {Schuell}, \citenamefont {Rabbel}, \citenamefont {Schmid},\ and\
  \citenamefont {Frey}}]{schuell2013polydispersity}%
  \BibitemOpen
  \bibfield  {author} {\bibinfo {author} {\bibfnamefont {Christoph}\
  \bibnamefont {Schuell}}, \bibinfo {author} {\bibfnamefont {Hauke}\
  \bibnamefont {Rabbel}}, \bibinfo {author} {\bibfnamefont {Friederike}\
  \bibnamefont {Schmid}}, \ and\ \bibinfo {author} {\bibfnamefont {Holger}\
  \bibnamefont {Frey}},\ }\bibfield  {title} {\enquote {\bibinfo {title}
  {Polydispersity and molecular weight distribution of hyperbranched graft
  copolymers via ``hypergrafting{''} of abm monomers from polydisperse
  macroinitiator cores: Theory meets synthesis},}\ }\href {\doibase
  10.1021/ma401119r} {\bibfield  {journal} {\bibinfo  {journal}
  {Macromolecules}\ }\textbf {\bibinfo {volume} {46}},\ \bibinfo {pages}
  {5823--5830} (\bibinfo {year} {2013})}\BibitemShut {NoStop}%
\bibitem [{\citenamefont {Schmid}(1998)}]{Schmid1998}%
  \BibitemOpen
  \bibfield  {author} {\bibinfo {author} {\bibfnamefont {F.}~\bibnamefont
  {Schmid}},\ }\bibfield  {title} {\enquote {\bibinfo {title}
  {{Self-consistent-field theories for complex fluids}},}\ }\href {\doibase
  10.1088/0953-8984/10/37/002} {\bibfield  {journal} {\bibinfo  {journal}
  {Journal of Physics Condensed Matter}\ }\textbf {\bibinfo {volume} {10}},\
  \bibinfo {pages} {8105--8138} (\bibinfo {year} {1998})}\BibitemShut {NoStop}%
\bibitem [{\citenamefont {Rabbel}\ \emph {et~al.}(2015)\citenamefont {Rabbel},
  \citenamefont {Frey},\ and\ \citenamefont {Schmid}}]{rabbel2013statistical}%
  \BibitemOpen
  \bibfield  {author} {\bibinfo {author} {\bibfnamefont {Hauke}\ \bibnamefont
  {Rabbel}}, \bibinfo {author} {\bibfnamefont {Holger}\ \bibnamefont {Frey}}, \
  and\ \bibinfo {author} {\bibfnamefont {Friederike}\ \bibnamefont {Schmid}},\
  }\bibfield  {title} {\enquote {\bibinfo {title} {Statistical properties of
  linear-hyperbranched graft copolymers prepared via ``hypergrafting{''} of abm
  monomers from linear b-functional core chains: A molecular dynamics
  simulation},}\ }\href {\doibase 10.1063/1.4935371} {\bibfield  {journal}
  {\bibinfo  {journal} {J. Chem. Phys.}\ }\textbf {\bibinfo {volume} {143}},\
  \bibinfo {pages} {243125} (\bibinfo {year} {2015})}\BibitemShut {NoStop}%
\bibitem [{\citenamefont {Anderson}\ \emph {et~al.}(2020)\citenamefont
  {Anderson}, \citenamefont {Glaser},\ and\ \citenamefont
  {Glotzer}}]{Anderson2020}%
  \BibitemOpen
  \bibfield  {author} {\bibinfo {author} {\bibfnamefont {Joshua~A.}\
  \bibnamefont {Anderson}}, \bibinfo {author} {\bibfnamefont {Jens}\
  \bibnamefont {Glaser}}, \ and\ \bibinfo {author} {\bibfnamefont {Sharon~C.}\
  \bibnamefont {Glotzer}},\ }\bibfield  {title} {\enquote {\bibinfo {title}
  {{HOOMD-blue: A Python package for high-performance molecular dynamics and
  hard particle Monte Carlo simulations}},}\ }\href {\doibase
  10.1016/J.COMMATSCI.2019.109363} {\bibfield  {journal} {\bibinfo  {journal}
  {Computational Materials Science}\ }\textbf {\bibinfo {volume} {173}},\
  \bibinfo {pages} {109363} (\bibinfo {year} {2020})},\ \Eprint
  {http://arxiv.org/abs/1308.5587} {1308.5587} \BibitemShut {NoStop}%
\bibitem [{\citenamefont {Zimm}(1948)}]{zimm1948apparatus}%
  \BibitemOpen
  \bibfield  {author} {\bibinfo {author} {\bibfnamefont {Bruno~H}\ \bibnamefont
  {Zimm}},\ }\bibfield  {title} {\enquote {\bibinfo {title} {Apparatus and
  methods for measurement and interpretation of the angular variation of light
  scattering; preliminary results on polystyrene solutions},}\ }\href {\doibase
  10.1063/1.1746740} {\bibfield  {journal} {\bibinfo  {journal} {The Journal of
  Chemical Physics}\ }\textbf {\bibinfo {volume} {16}},\ \bibinfo {pages}
  {1099--1116} (\bibinfo {year} {1948})}\BibitemShut {NoStop}%
\bibitem [{\citenamefont {Sanchez}\ and\ \citenamefont
  {Lacombe}(1978)}]{sanchez-lacombe}%
  \BibitemOpen
  \bibfield  {author} {\bibinfo {author} {\bibfnamefont {Isaac~C.}\
  \bibnamefont {Sanchez}}\ and\ \bibinfo {author} {\bibfnamefont {Robert~H.}\
  \bibnamefont {Lacombe}},\ }\bibfield  {title} {\enquote {\bibinfo {title}
  {Statistical thermodynamics of polymer solutions},}\ }\href {\doibase
  10.1021/ma60066a017} {\bibfield  {journal} {\bibinfo  {journal}
  {Macromolecules}\ }\textbf {\bibinfo {volume} {11}},\ \bibinfo {pages}
  {1145--1156} (\bibinfo {year} {1978})}\BibitemShut {NoStop}%
\bibitem [{\citenamefont {Grason}\ and\ \citenamefont
  {Kamien}(2005)}]{Grason2005}%
  \BibitemOpen
  \bibfield  {author} {\bibinfo {author} {\bibfnamefont {Gregory~M.}\
  \bibnamefont {Grason}}\ and\ \bibinfo {author} {\bibfnamefont {Randall~D.}\
  \bibnamefont {Kamien}},\ }\bibfield  {title} {\enquote {\bibinfo {title}
  {Self-consistent field theory of multiply branched block copolymer melts},}\
  }\href {\doibase 10.1103/PhysRevE.71.051801} {\bibfield  {journal} {\bibinfo
  {journal} {Phys. Rev. E}\ }\textbf {\bibinfo {volume} {71}},\ \bibinfo
  {pages} {051801} (\bibinfo {year} {2005})}\BibitemShut {NoStop}%
\bibitem [{\citenamefont {Müller}\ and\ \citenamefont
  {Schmid}(2005)}]{muller_fschmid}%
  \BibitemOpen
  \bibfield  {author} {\bibinfo {author} {\bibfnamefont {Marcus}\ \bibnamefont
  {Müller}}\ and\ \bibinfo {author} {\bibfnamefont {Friederike}\ \bibnamefont
  {Schmid}},\ }\bibfield  {title} {\enquote {\bibinfo {title} {Incorporating
  fluctuations and dynamics in self-consistent field theories for polymer
  blends},}\ }in\ \href {\doibase 10.1007/b136794} {\emph {\bibinfo {booktitle}
  {Advanced Computer Simulation Approaches for Soft Matter Sciences II}}}\
  (\bibinfo  {publisher} {Springer},\ \bibinfo {year} {2005})\ pp.\ \bibinfo
  {pages} {1--58}\BibitemShut {NoStop}%
\bibitem [{\citenamefont {Yong}\ and\ \citenamefont {Kim}()}]{yong2025dynamic}%
  \BibitemOpen
  \bibfield  {author} {\bibinfo {author} {\bibfnamefont {Daeseong}\
  \bibnamefont {Yong}}\ and\ \bibinfo {author} {\bibfnamefont {Jaeup~U.}\
  \bibnamefont {Kim}},\ }\bibfield  {title} {\enquote {\bibinfo {title}
  {Dynamic programming for chain propagator computation of branched block
  copolymers in polymer field theory simulations},}\ }\href {\doibase
  10.1021/acs.jctc.5c00103} {\bibfield  {journal} {\bibinfo  {journal} {Journal
  of Chemical Theory and Computation}\ }\textbf {\bibinfo {volume} {21}},\
  \bibinfo {pages} {3676--3690}}\BibitemShut {NoStop}%
\bibitem [{\citenamefont {Frey}\ and\ \citenamefont
  {H{\"o}lter}(1999)}]{frey1999degree}%
  \BibitemOpen
  \bibfield  {author} {\bibinfo {author} {\bibfnamefont {H}~\bibnamefont
  {Frey}}\ and\ \bibinfo {author} {\bibfnamefont {D}~\bibnamefont
  {H{\"o}lter}},\ }\bibfield  {title} {\enquote {\bibinfo {title} {Degree of
  branching in hyperbranched polymers. 3 copolymerization of abm-monomers with
  ab and abn-monomers},}\ }\href {\doibase
  https://doi.org/10.1002/(SICI)1521-4044(19990201)50:2/3<67::AID-APOL67>3.0.CO;2-W}
  {\bibfield  {journal} {\bibinfo  {journal} {Acta polymerica}\ }\textbf
  {\bibinfo {volume} {50}},\ \bibinfo {pages} {67--76} (\bibinfo {year}
  {1999})}\BibitemShut {NoStop}%
\bibitem [{\citenamefont {Hanselmann}\ \emph {et~al.}(1998)\citenamefont
  {Hanselmann}, \citenamefont {H{\"{o}}lter}, \citenamefont {Frey},
  \citenamefont {Hanselmann}, \citenamefont {H{\"{o}}lter},\ and\ \citenamefont
  {Frey}}]{Hanselmann1998}%
  \BibitemOpen
  \bibfield  {author} {\bibinfo {author} {\bibfnamefont {Ralf}\ \bibnamefont
  {Hanselmann}}, \bibinfo {author} {\bibfnamefont {Dirk}\ \bibnamefont
  {H{\"{o}}lter}}, \bibinfo {author} {\bibfnamefont {Holger}\ \bibnamefont
  {Frey}}, \bibinfo {author} {\bibfnamefont {Ralf}\ \bibnamefont {Hanselmann}},
  \bibinfo {author} {\bibfnamefont {Dirk}\ \bibnamefont {H{\"{o}}lter}}, \ and\
  \bibinfo {author} {\bibfnamefont {Holger}\ \bibnamefont {Frey}},\ }\bibfield
  {title} {\enquote {\bibinfo {title} {{Hyperbranched Polymers Prepared via the
  Core-Dilution/Slow Addition Technique: Computer Simulation of Molecular
  Weight Distribution and Degree of Branching}},}\ }\href {\doibase
  10.1021/MA971197R} {\bibfield  {journal} {\bibinfo  {journal} {MaMol}\
  }\textbf {\bibinfo {volume} {31}},\ \bibinfo {pages} {3790--3801} (\bibinfo
  {year} {1998})}\BibitemShut {NoStop}%
\bibitem [{\citenamefont {Israelachvili}\ \emph {et~al.}(1976)\citenamefont
  {Israelachvili}, \citenamefont {Mitchell},\ and\ \citenamefont
  {Ninham}}]{israelachvili1976theory}%
  \BibitemOpen
  \bibfield  {author} {\bibinfo {author} {\bibfnamefont {Jacob~N}\ \bibnamefont
  {Israelachvili}}, \bibinfo {author} {\bibfnamefont {D~John}\ \bibnamefont
  {Mitchell}}, \ and\ \bibinfo {author} {\bibfnamefont {Barry~W}\ \bibnamefont
  {Ninham}},\ }\bibfield  {title} {\enquote {\bibinfo {title} {Theory of
  self-assembly of hydrocarbon amphiphiles into micelles and bilayers},}\
  }\href {\doibase https://doi.org/10.1039/F29767201525} {\bibfield  {journal}
  {\bibinfo  {journal} {Journal of the Chemical Society, Faraday Transactions
  2: Molecular and Chemical Physics}\ }\textbf {\bibinfo {volume} {72}},\
  \bibinfo {pages} {1525--1568} (\bibinfo {year} {1976})}\BibitemShut {NoStop}%
\bibitem [{\citenamefont {Hansen}(2007)}]{Hansen_solubility}%
  \BibitemOpen
  \bibfield  {author} {\bibinfo {author} {\bibfnamefont {Charles~M.}\
  \bibnamefont {Hansen}},\ }\href@noop {} {\emph {\bibinfo {title} {{Hansen's
  Solubility Parameters: A User's Handbook}}}},\ \bibinfo {edition} {2nd}\ ed.\
  (\bibinfo  {publisher} {CRC Press, Taylor \& Francis group},\ \bibinfo {year}
  {2007})\BibitemShut {NoStop}%
\bibitem [{\citenamefont {Zhang}\ \emph {et~al.}(1996)\citenamefont {Zhang},
  \citenamefont {Jackson},\ and\ \citenamefont {Burt}}]{zhang1996development}%
  \BibitemOpen
  \bibfield  {author} {\bibinfo {author} {\bibfnamefont {Xichen}\ \bibnamefont
  {Zhang}}, \bibinfo {author} {\bibfnamefont {John~K}\ \bibnamefont {Jackson}},
  \ and\ \bibinfo {author} {\bibfnamefont {Helen~M}\ \bibnamefont {Burt}},\
  }\bibfield  {title} {\enquote {\bibinfo {title} {Development of amphiphilic
  diblock copolymers as micellar carriers of taxol},}\ }\href {\doibase
  https://doi.org/10.1016/0378-5173(95)04386-1} {\bibfield  {journal} {\bibinfo
   {journal} {International journal of pharmaceutics}\ }\textbf {\bibinfo
  {volume} {132}},\ \bibinfo {pages} {195--206} (\bibinfo {year}
  {1996})}\BibitemShut {NoStop}%
\bibitem [{\citenamefont {Zhu}\ \emph {et~al.}(2011)\citenamefont {Zhu},
  \citenamefont {Zhou},\ and\ \citenamefont {Yan}}]{Zhu2011}%
  \BibitemOpen
  \bibfield  {author} {\bibinfo {author} {\bibfnamefont {Xinyuan}\ \bibnamefont
  {Zhu}}, \bibinfo {author} {\bibfnamefont {Yongfeng}\ \bibnamefont {Zhou}}, \
  and\ \bibinfo {author} {\bibfnamefont {Deyue}\ \bibnamefont {Yan}},\
  }\bibfield  {title} {\enquote {\bibinfo {title} {Influence of branching
  architecture on polymer properties},}\ }\href {\doibase 10.1002/polb.22320}
  {\bibfield  {journal} {\bibinfo  {journal} {J. Pol. Science Part B -- Polymer
  Physics}\ }\textbf {\bibinfo {volume} {49}},\ \bibinfo {pages} {1277--1286}
  (\bibinfo {year} {2011})}\BibitemShut {NoStop}%
\bibitem [{\citenamefont {Qiao}\ \emph {et~al.}(2024)\citenamefont {Qiao},
  \citenamefont {Giannakou},\ and\ \citenamefont {Schmid}}]{qiao}%
  \BibitemOpen
  \bibfield  {author} {\bibinfo {author} {\bibfnamefont {Le}~\bibnamefont
  {Qiao}}, \bibinfo {author} {\bibfnamefont {Marios}\ \bibnamefont
  {Giannakou}}, \ and\ \bibinfo {author} {\bibfnamefont {Friederike}\
  \bibnamefont {Schmid}},\ }\bibfield  {title} {\enquote {\bibinfo {title} {An
  efficient and accurate scf algorithm for block copolymer films and brushes
  using adaptive discretizations},}\ }\href {\doibase 10.3390/polym16091228}
  {\bibfield  {journal} {\bibinfo  {journal} {Polymers}\ }\textbf {\bibinfo
  {volume} {16}} (\bibinfo {year} {2024}),\ 10.3390/polym16091228}\BibitemShut
  {NoStop}%
\bibitem [{\citenamefont {Wang}\ and\ \citenamefont {Wang}(2012)}]{Wang2012}%
  \BibitemOpen
  \bibfield  {author} {\bibinfo {author} {\bibfnamefont {Rui}\ \bibnamefont
  {Wang}}\ and\ \bibinfo {author} {\bibfnamefont {Zhen-Gang}\ \bibnamefont
  {Wang}},\ }\bibfield  {title} {\enquote {\bibinfo {title} {Theory of polymers
  in poor solvent: Phase equilibrium and nucleation behavior},}\ }\href
  {\doibase https://doi.org/10.1021/ma301049m} {\bibfield  {journal} {\bibinfo
  {journal} {Macromolecules}\ }\textbf {\bibinfo {volume} {45}},\ \bibinfo
  {pages} {6266--6271} (\bibinfo {year} {2012})}\BibitemShut {NoStop}%
\end{thebibliography}%

\clearpage

\renewcommand{\thetable}{S\arabic{table}}
\renewcommand{\thefigure}{S\arabic{figure}}
\renewcommand{\thesection}{S\arabic{section}}
\renewcommand{\theHtable}{S\arabic{table}}
\renewcommand{\theHfigure}{S\arabic{figure}}
\renewcommand{\theHsection}{S\arabic{section}}
\setcounter{figure}{0}
\setcounter{section}{0}
\section*{Supporting Information}
\begin{figure}
    \centering
    \includegraphics[width=0.48\textwidth]{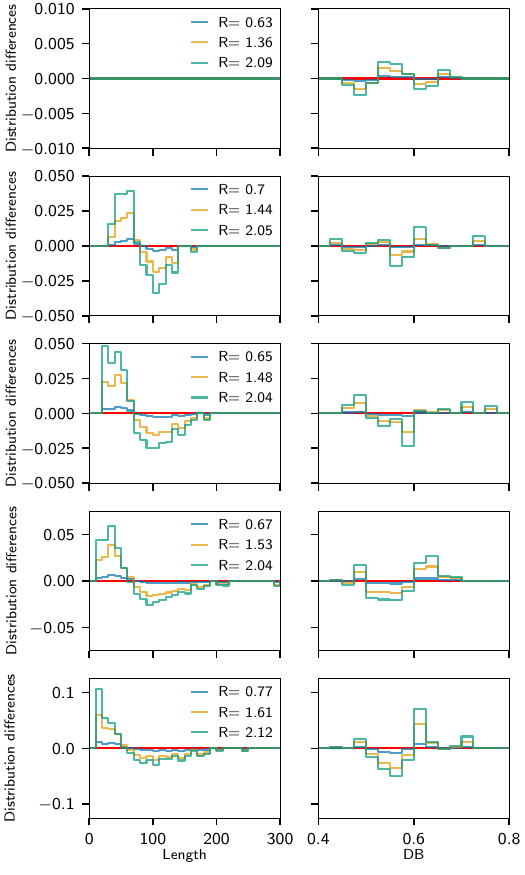}
    \caption{Probability distributions differences between the micelle and the bath, in the length (left) and degree of branching (DB) (right) for the hyperbranched part of the LHBCs for different values of PDI. From top to bottom each row corresponds to PDIs of $1,\, 1.1,\, 1.2, \,1.3$ and $1.4$. The different colors correspond to the critical micelle (micelle at maximum), the equilibrium micelle (micelle at minimum) and the largest micelle we obtained from Fig.~\ref{fig:size_dis} in the main text respectively.}
    \label{fig:distributions_shift}
\end{figure}

\begin{figure}
    \centering
    \includegraphics[width=0.48\textwidth]{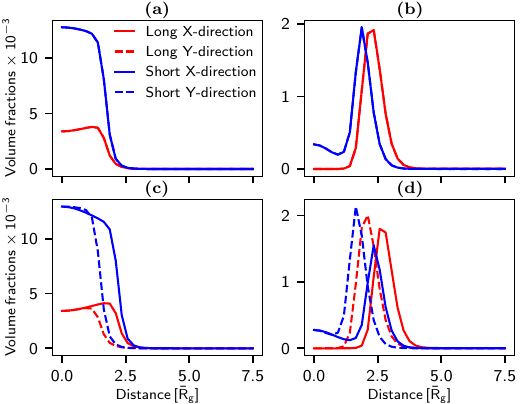}
    \caption{(a) Solvophobic and (b) solvophilic volume fraction contribution of a short chain (blue) and a long chain (red) in a spherical micelle. (c) Solvophobic and (d) solvophilic volume fraction contribution of a short chain (blue) and a long chain (red) in a micelle with an asphericity of A=$0.8$ in the long axis direction X (solid line) and the short axis direction Y (dashed line). The short chain has a solvophilic length of $23$ while the long chain has a length of $109$, while the solvophobic block is $84$ for both as stated in the main text. The results correspond to the system with PDI = $1.4$ as shown in Fig.~\ref{fig:asphericities} in the main text. }
    \label{fig:densxy}
\end{figure}

Here we show some additional results to further illustrate or support arguments made in the main text.
Fig.~ \ref{fig:distributions_shift} shows the distribution differences between micelles of different sizes and the bath. As is evident, smaller chains contribute to an increase in the size of the micelle, although from the PDI = $1$ case it is also clear that also chains with certain degrees of branching might be more favored with increasing micelle size. The latter effect is albeit less pronounced.
In Fig.~ \ref{fig:densxy}, we plot the volume fraction contributions from two different types of polymers, a short and a long one, in a LHBC system containing an equilibrium spherical micelle (Fig.\ \ref{fig:densxy}a,b) and a stretched micelle (Fig.\ \ref{fig:densxy}c,d), as obtained from the work presented in the main text. To interpret the results, we recall some specifics of our setup: Whereas the sets of chain types in an SCF calculation are constructed such that the molecular weights distributions roughly follow a Schulz-Zimm distribution, all chain types within such a set have equal {\em a priori} probability. Moreover, the solvophobic blocks of all chains have equal length. Therefore, Figs.\ \ref{fig:densxy}a,c illustrate that micelles preferably recruit chains with short solvophilic parts, consistent with Fig.\ \ref{fig:distributions_shift} and Fig. \ref{fig:cmcs}c in the main text. Interestingly, however, the contribution of solvophilic monomers to the total volume fraction is comparable for short and long chain types, the main difference being that the solvophilic monomers of shorter chains tend to be closer to the core than those of longer chains.
 Additionally, the short chain's solvophilic contribution shown in (b) exhibits two maxima, one in the center and one near the corona, which is not seen at all for the long chain. As explained in the main text, this effect is attributed to the fact that the short chain is almost completely solvophobic, such that solvophilic blocks can be pulled inside the core at low energy cost. The same trends are observed in the aspherical micelle shown in (c) and (d). In the aspherical case, the main maxima of the solvophilic distributions shift to larger distances along the long axis and smaller distances along the short axis for both short and long chains, while their positions relative to each other remain unchanged.
 
\begin{figure}
    \centering
    \includegraphics[width=0.48\textwidth]{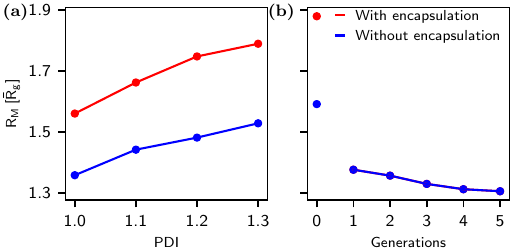}
    \caption{Radius $R_M$ of the micelle as a function of the
    polydispersity for the LHBC systems (a) and as a function of the
    number of generations for the LDBC systems (b). The red lines
    correspond to the case where the calculations have been
    performed with encapsulation of a homopolymer and blue without.
    All calculations were performed for the equilibrium spherical
    micelle. Note that the red and blue lines coincide in (b).\label{fig:radiiencaps}}
\end{figure}

Fig.\ \ref{fig:radiiencaps} shows the radius of the micelles for
spherical micelles with and without encapsulation of homopolymers.
The results demonstrate that the radius of the micelle increases with
encapsulation for all LHBC systems, whereas for the LDBC systems, the radius
of the micelles is unchanged. This is consistent with the picture that
LDBC micelles are less prone to changes in their structure and therefore
accommodate a reduced payload compared to the other classes of polymers.

\begin{figure}[H]
    \centering
    \includegraphics[width=0.48\textwidth]{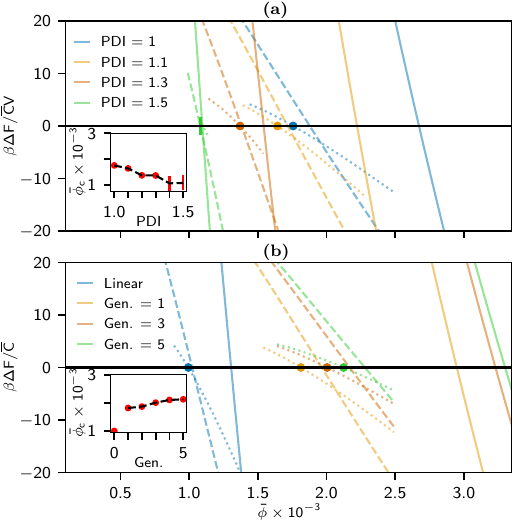}
    \caption{Rescaled free energy difference $\Delta F$ between
    homogeneous and inhomogeneous states in the presence of
    homopolymer model drug ($\Bar{\phi}_h = 10^{-5}$ in the bath) against
    the average copolymer volume fraction $\Bar{\phi}$ in the bath for
    polydisperse LHBCs and monodisperse linear and LDBCs respectively.
    Each color corresponds to a different polymer system, while the
    style of the line corresponds to lamella (solid line), cylindrical
    (dashed line) and spherical (dotted line) micelle states. The
    inset axes in figures (a) and (b) show the state with the lowest
    critical concentration $\Bar{\phi}_c$ against PDI and number of
    generations respectively. The circular \mbox{(\protect\clcir)} and
    rectangular \mbox{(\protect\rect)} symbols indicate spherical micelles
    and cylindrical micelles respectively as the equilibrium morphology. These symbols are also
    shown in Fig.~\ref{fig:cmcs} in the main text. Note that ${\Bar{\phi}_c}$
    and ${\Bar{\phi}}$ are calculated without ${\Bar{\phi}_h}.$}
    \label{fig:cmcsencaps}
\end{figure}

In Fig.\ \ref{fig:cmcsencaps}, the free energy difference between the
inhomogeneous state and the homogeneous state, $\mathrm{\Delta F}$, is
shown for the different polymer systems with encapsulated
homopolymers. Also shown is the the critical volume fraction
${\bar{\phi}_{c}}$ and the equilibrium morphology.

\begin{figure}[H]
    \centering
    \includegraphics[width=0.48\textwidth]{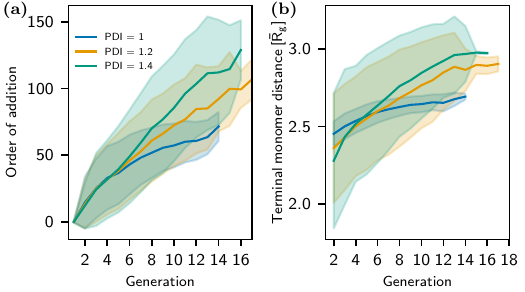}
    \caption{(a) Order of addition of
    monomers vs. their
    corresponding generation in each polymer molecule.
    Results from BA batch. (b)
    Mean distance of terminal monomers from the center of the
    equilibrium micelle vs.\  corresponding generation. Results from averaging
    B$1$-B$4$.} 
    \label{fig:order_gen}
\end{figure}

Finally, as illustrated in Figs.\
\ref{fig:order_gen}a and \ref{fig:order_gen}b, there 
exists a positive correlation between the order of addition of
the $\mathrm{AB_{2}}$ monomers, the generation they occupy on the
growing molecule, and their distance from the micelle center. Although
the spread is significant, adding suitably functionalized
$\mathrm{AB_{2}}$ monomers either at the very beginning
or the very end of the synthesis process could be viable
strategies for positioning functional units inside or at the outer
end of the corona, respectively.
\end{document}